\documentclass[11pt,amsmath,amssymb,aps,nofootinbib,prd,preprint,superscriptaddress]{revtex4-2}

\usepackage[normalem]{ulem}

\usepackage{tikz}
\usetikzlibrary{arrows.meta, positioning}

\usepackage{hyperref}
\usepackage{comment}
\usepackage{color}

\usepackage{setspace}
\usepackage{cancel}
\usepackage{graphicx}
\usepackage{enumerate} 

\tikzset{myarrow/.style={->, shorten >=5pt, shorten <=5pt, >=Latex,thick}}

\def\be{\begin{equation}}
\def\ee{\end{equation}}
\def\ba{\begin{eqnarray}}
\def\ea{\end{eqnarray}}
\def\bal{\begin{equation} \begin{aligned}}
\def\eal{\end{aligned}\end{equation} }

\def\bi{\begin{itemize}}
\def\ei{\end{itemize}}

\def\zb{\bar{z}}

\def\w{\omega}

\def\nh{\hat{n}}

\def\ph{\hat{p}}
\def\I{\mathcal{I}}

\def\w{\omega}

\newcommand{\ov}[2]{\overset{\scriptscriptstyle #1}{#2}\vphantom{#1}}

\newcommand{\atscri}[1]{\big|_{\scri^{#1}}}

\def\harm{\text{harm}}

\def\scri{\mathcal{I}}

\def\htilde{\tilde{h}}
\def\div{\text{div}}
\def\rad{\text{rad}}

\def\inn{\text{in}}
\def\out{\text{out}}
\def\M{\mathcal{M}}
\def\O{\mathcal{O}}

\def\N{\mathcal{N}}
\def\V{\mathcal{V}}
\def\D{\mathcal{D}}

\def\P{\mathcal{P}}
\def\Q{\mathcal{Q}}
\def\Nt{\tilde{\mathcal{N}}}

\def\B{\mathcal{B}}

\def\quadr{\text{quad}}

\def\mbb{\bar{\text{m}}}
\def\mb{\bar{m}}
\def\m{\text{m}}
\def\st{\tilde{s}}
\def\pt{\tilde{p}}
\def\ellb{\bar{\ell}}

\def\p{\text{p}}
\def\Chat{\widehat{C}}
\def\ghat{\widehat{g}}

\begin{document}

\title{Peeling-violating coefficients in classical gravitational scattering}

\author{Gianni Boschetti}
\email{gboschetti@fing.edu.uy}
\affiliation{Instituto de Física, Facultad  de  Ingeniería, Universidad  de  la  Rep\'ublica \\
Julio Herrera y Reissig 565,  Montevideo,  Uruguay}

\author{Miguel Campiglia}
\email{miguel.campiglia@fcien.edu.uy}
\affiliation{Instituto de Física, Facultad  de  Ciencias, Universidad  de  la  Rep\'ublica \\
Ig\'ua  4225,  Montevideo,  Uruguay}

\begin{abstract}
Using matching properties of the gravitational field at timelike and spatial infinity, together with  universal formulas for gravitational wave tails, 
we obtain exact expressions for the peeling-violating components of the Weyl tensor at future and past null infinity. The coefficients depend solely on incoming scattering data and reduce, in the Newtonian limit, to those found by Damour long ago. The computation is facilitated by the observation that the tail and peeling-violating coefficients organize into a ``celestial diamond'' introduced in the context of celestial holography. The analysis suggests a new matching condition at spatial infinity which, together with previously found matching at timelike infinity,  capture both  tail and peeling-violation formulas. We conclude by  pointing out that stronger peeling violations  occur in perturbative quantum gravity, 
 in agreement  with a recently reported amplitude-based result.
\end{abstract}

\maketitle

\newpage

\tableofcontents

\section{Introduction}

In asymptotically flat spacetimes, the gravitational field can exhibit two types  of “slow” decay near null infinity: in the radial distance $r$ and in (retarded or advanced) time $u$.  The foundational  works 
\cite{sachsoutgoing,bondi,sachs,NPspincoeff,penrosezrm}   excluded the first possibility by imposing the so-called peeling property. It was later recognized that this requirement is too restrictive \cite{bardeenpress,cauchtorrence,novakgold,porrill,winicour} and  incompatible with gravitational scattering \cite{walkerwilll,damour}. Since then, the subject of \emph{partially peeling} spacetimes has been developed   \cite{chrusciel,kroon1,kroon2,kroonout,kroon3,christo,friedrichstationary,friedrich} and remains a topic of current interest \cite{kehrberger1,kehrberger2,kehrberger3,kehrberger4,gajic,caponedg4,geillerpeeling,godazgar,marajh,bieri,schneiderkhera,radu,briceno,compererobert2}.

The second type of slow decay refers to \emph{tails}, generically arising in gravitational waves due to non-linearities and curvature distortions \cite{bonnor,ctnj,DeWittBrehme,peters,price}.
The effect has been extensively studied in the context of gravitational radiation from compact binaries (see \cite{hereditary,blanchetschaefer,blanchetlrr,trestini} and references therein), where it leads to observable  contributions to the predicted waveforms \cite{blanchetsathyaprl,blanchetsathyacqg}.

Despite their apparently different character, 
it has been repeatedly suggested \cite{kroon3,christo,friedrichstationary,friedrich,kehrberger1,kehrberger2,kehrberger3,kehrberger4,gajic,geillerpeeling,radu,briceno,compererobert2,marajh} that the two types of slow decay  have a common origin. Inspired by these considerations, in this paper we derive, in the context of scattering spacetimes,   peeling-violating coefficients from $O(1/u)$-tail formulas and matching properties of the gravitational field. Our analysis further identifies  a new set of matching conditions at spatial infinity as the common origin of both phenomena.

We follow the perspective on asymptotically flat gravity initiated by Strominger and collaborators  \cite{stromgravscatt,stromST,stromvirasoro}, which brings to the forefront the role of \emph{soft theorems} \cite{weinberg,grossjackiw,white,stromingercachazo}. Of direct relevance to our analysis  are: i) the identification of soft theorems as matching conditions at spatial infinity \cite{stromgravscatt,stromST}; ii) their reformulation in the language of conformal field theory (CFT)  \cite{stromvirasoro,stromym,2dkacmoody,2dstress}; and iii) the soft-theorem-based approach to gravitational wave tails developed by Laddha, Saha, Sahoo, and Sen (LSSS) \cite{laddhasen1,laddhasen,sahoosen,proofdeq4}.

Our starting point is the LSSS description of tails. This approach bypasses the difficult problem of solving the evolution of the radiating system -- which is at the core of traditional treatments -- by allowing the tail coefficients to depend on initial \emph{and} final data. This results in universal identities \cite{proofdeq4,sahoosubsub,rewritten,ghoshsahoo,senreview} that can be used as consistency checks for calculations of gravitational wave emission \cite{ciafcolfven,addazibianchiven,manughoshladdhaathira,vishwakarma,bautiladdha,spectra,georgoheissrusso,georgogoncheissjpm,alessiovddgonzorosi,alessiopdv,Bini:2024rsy}. Here we focus on the leading tail, capturing the $1/u$ component of the waveform. In frequency space this leads to a $\log \w$ term in the soft expansion, with coefficient given by the \emph{difference} of the early- and late-time tail coefficients.

In \cite{gianni2} we showed that the logarithmic soft theorem follows from matching conditions at timelike and spatial infinity. The analysis of \cite{gianni2} left undetermined peeling-violating coefficients, which we now fix by bringing in the LSSS formula for the \emph{individual} tail components. In what appears to be an unexpected cancellation, the resulting expressions turn out to depend solely on incoming data.\footnote{This is consistent with a theorem due to Blanchet \cite{blanchetpeeling}, which establishes peeling assuming past stationarity—a condition violated by incoming scattering data \cite{damour,blanchetpeeling}.} We interpret this result as originating from a new matching condition at spatial infinity which, together with those found in \cite{gianni2}, implies both  tail and peeling violation formulas.

We  adopt Bondi-Sachs gauge fixing conditions to describe the metric near null infinity and use  stereographic coordinates $(z,\zb)$ for the angular directions. In this setting, the peeling-violating coefficients are given by a $O(r^0)$ traceless 2d tensor $p_{zz}$ and a $O(\ln r)$ angular momentum aspect $ \ov{\log r}{\N_z}$ that are related by \cite{sachs,winicour}
\be \label{divPintro}
  D^z {p_{zz}} = G \, \ov{\log r}{\N_z},
\ee
where $G$ is Newton constant.   It is the vector $ \ov{\log r}{\N_z}$ that  gets  determined from the matching and tail formulas.  To solve for  $p_{zz}$ in \eqref{divPintro} we take advantage of the ``celestial diamond'' picture of \cite{diamond1,diamond2}. As already noted in \cite{2dstress}, $p_{zz}$  has the  conformal dimension of a  2d stress tensor. This lets us identify \eqref{divPintro} as part of a larger set of fields related by derivatives according to the  diagram \cite{diamond1},
\be  \label{diamondintro}
\begin{tikzpicture}[baseline={(current bounding box.center)}]
  \def\L{1.3}  
  \def\l{0.9}  

  \node (A) at (0,\L) {$\V^z$};
  \node (B) at (2*\L, -\L) {$p_{zz}$};
  \node (C) at (\L,-2*\L) {$G \ov{\log r}{\N_z}$};
  \node (D) at (-\L,0) {$\pt^{zz}$};

\node at (-\l,\l) {$D^z$};
\node at (-\l+2*\L,\l-2*\L) {$D^z$};
\node at (\l+\L/2,\l-\L/2) {$D^3_z$};
\node at (\l-\L/2,\l-3*\L/2) {$D^3_z$};

\draw[->] (A) --  (B);
  \draw[->] (B) --  (C);
  \draw[->] (A) --  (D);
  \draw[->] (D) --  (C);

\end{tikzpicture} \quad .
\ee
The extra fields are the ``shadow transformed'' $\pt^{zz}$    \cite{kapecmitra,donnaypuhmstrom,donnaypastepuhm}  and the  ``reparametrization mode''  $\V^z$ \cite{repmode,srmode,pastebaner} associated to the would-be  stress tensor.  The first one  is related to the $1/u$ tails  while the second one lacks a direct spacetime interpretation. It however 
allows  us to find simple and closed formula for the solution to Eq. \eqref{divPintro}. As an illustration, at future null infinity we find
\be \label{Vscriplus}
{\V^z}\atscri{+} = 4 G^2 \partial^z n_{[\mu}  n_{\nu]} P^\mu \sum_{i \in \inn} \log( p_i \cdot n) p^{\nu}_i,
\ee
where $n^\mu$ is a future-directed null 4-vector associated to the angular direction;   $\{p_i^\mu, i \in \inn\}$ is the set of incoming momenta; and $P^\mu$ the total momentum.\footnote{We adopt mostly plus signature conventions, and include an overall minus sign for incoming momenta, so that $ p_i \cdot n>0 $ in \eqref{Vscriplus}.} Applying $D_z^3$ to \eqref{Vscriplus} yields the future peeling-violating coefficient $p_{zz}\atscri{+}$. In the Newtonian limit, the resulting expression reduces to the one obtained by Damour  \cite{damour} through a perturbative computation. Similar considerations apply for the past null infinity coefficients.

In terms of the often-used Weyl scalars  \cite{NPspincoeff}, the coefficients \eqref{divPintro} lead to a violation of peeling in $\Psi_0=O(r^{-4})$ and  $\Psi_1=O(r^{-4} \log r)$. Intriguingly,  a recent amplitude-based analysis \cite{radu} has led the authors to conclude that $\Psi_0=O(r^{-4})$ and  $\Psi_1=O(r^{-3} )$.  Their result is in tension  with a ``partial peeling'' theorem due to Valiente Kroon \cite{kroon1,kroon2}, which fixes the decay of the $\Psi_{k>0}$ scalars once $\Psi_0$ is specified. We will argue that the decay in $\Psi_1$ found in \cite{radu} is indeed expected in perturbative quantum gravity, but that it should be accompanied by a $\Psi_0=O(r^{-3})$. The mechanism behind this  quantum peeling violation is directly related to the quantum terms in the logarithmic soft theorem \cite{sahoosen}, whose  vanishing mechanism in the classical limit is notoriously subtle \cite{manughoshladdhaathira,vishwakarma}.

\subsection*{Outline}
In section \ref{tailsec} we review the $O(1/u)$ tail formula from the perspective presented in \cite{gianni1}, which highlights the role of  \emph{logarithmic translations} \cite{Bergmann:1961zz,aalog}.
 In section \ref{bondipartialpeelsec} we review the Bondi metric with ``minimal'' peeling-violating terms, as first considered by Winicour \cite{winicour}.  In section \ref{cftsec} we review the celestial CFT perspective on metric coefficients at null infinity and introduce a set of four elementary ``soft tensors''  underlying the diamond diagram \eqref{diamondintro}.  In section  \ref{peelcoefsec} we combine the   tail formulas,  the matchings of \cite{gianni2}, and the soft tensors to  obtain   formulas for the peeling-violating coefficients \eqref{divPintro}.  In section \ref{newmatching} these results are interpreted as originating from a set of matching conditions that includes a novel condition at spatial infinity. 
 Section \ref{quantumsec} provides a brief argument for why, quantum mechanically, stronger peeling violations are expected.  We conclude in section \ref{discussionsec} with a brief summary of results and open questions.

 The paper is complemented with three appendices:
  Appendix \ref{NPapp} reviews the definition of Weyl scalars, their (partial) peeling  and their relation with the coefficients \eqref{divPintro}. Appendix  \ref{damourapp} reviews Damour's peeling violation results  and translate them to the notation of the present paper. Appendix \ref{polyapp} reviews the classification of non-peeling decays due  to Valiente Kroon.

\section{Gravitational wave tails} \label{tailsec}

Let us denote by
\be \label{defhout}
h_{\mu \nu}\atscri{+}(u,\nh) := \lim_{r \to \infty} r (  g_{\mu \nu}-\eta_{\mu \nu}),
\ee
the gravitational waveform at future null infinity, where $r$ is a radial distance, $u$ a retarded time and $\nh$ the direction on the celestial sphere.  At asymptotic early/late times it behaves as \cite{senreview}
\be \label{htails}
h_{\mu \nu}\atscri{+}(u,\nh) \stackrel{u \to \pm \infty}{=} \ov{0}{h}^\pm_{\mu \nu}(\nh) + \frac{1}{u} \ov{1}{h}^\pm_{\mu \nu}(\nh) + \cdots,
\ee
where $\ov{0}{h}^\pm_{\mu \nu}$ capture the ``displacement'' or ``memory'' effect (see \cite{zhibo} and references therein) and $\ov{1}{h}^\pm_{\mu \nu}$ the leading tail. The dots include subleading tails decaying faster than $1/u$ \cite{senreview}.

Fourier transforming \eqref{htails} leads to a low-frequency or \emph{soft} expansion \cite{senreview}
\begin{equation} \label{hmunuomega}
\tilde{h}_{\mu\nu}\atscri{+}(\omega, \hat{n}) \stackrel{\w \to 0}{=}   \omega^{-1} \htilde^{(0)}_{\mu\nu}(\hat{n}) + \log \omega \, \htilde^{(\log)}_{\mu\nu}(\hat{n}) + \cdots 
\end{equation}
with\footnote{By giving the soft frequency a small positive/negative imaginary part, one can extract individual tail terms --not just their differences-- from the soft expansion. As in \cite{gianni1,gianni2}, we refer to the soft expansion/soft theorems to the limit with no such imaginary deformation.} 
\ba
\htilde^{(0)}_{\mu\nu}  &=&  i \Big(\ov{0}{h}^+_{\mu \nu}- \ov{0}{h}^-_{\mu \nu}\Big)  , \label{delh0eqhtilde0} \\
   \htilde^{(\log)}_{\mu\nu} &=& -\Big(\ov{1}{h}^+_{\mu \nu}- \ov{1}{h}^-_{\mu \nu}\Big).
\ea

The individual coefficients for the leading term in \eqref{htails} depend on the supertranslation frame implicitly assumed in \eqref{defhout}, but their difference is insensitive to this choice and given by \cite{BraginskyThorne,weinberg}
\be \label{diffh0}
\ov{0}{h}^+_{\mu \nu}- \ov{0}{h}^-_{\mu \nu} =-  4 G  \sum_i \frac{p^i_\mu p^i_\nu}{p_i \cdot n},
\ee
where  $n^\mu=(1,\nh)$ is the future directed null vector associated to $\nh$  and the sum runs over both incoming and outgoing particle momenta.\footnote{Here “particle” is used in a broad sense \cite{senreview} to denote any asymptotic constituent, including macroscopic bodies and radiation.  In particular, \eqref{diffh0} includes   gravitational radiation contributions \cite{thornenl}.}

The  tail coefficients in   \eqref{htails} are independent of  frame choices and admit universal formulas 
we review below. The formulas involve double particle sums, but for our purposes it will be convenient to express them as single sums with momentum-dependent coefficients \cite{laddhasen1}. The coefficients are constructed by momentum-dependent vectors $c_i^\mu$  that capture the logarithmic deviation of the  particles' asymptotic trajectories, according to
\be \label{asymtraj}
X^\mu_i(s_i)  \stackrel{s_i \to \pm \infty}{=}  s_i V^{\mu}_i  +  \log  |s_i| c^\mu_i +\cdots,
\ee
where $s_i$ is an affine parameter and the plus/minus infinity limit refers to outgoing/incoming.  As explained in \cite{gianni1}, these vectors are  defined modulo global shifts due to \emph{logarithimic translations} \cite{Bergmann:1961zz,aalog},
\be \label{delLci}
\delta_L c_i^\mu = - L^\mu,
\ee
where  $L^\mu$ is an arbitrary constant vector.  In the analysis of \cite{laddhasen1,laddhasen,sahoosen,proofdeq4} this freedom is frozen by the harmonic gauge condition, in which case the vectors take the form \cite{sahoosen}
\ba \label{charm}
c^{\harm \, \mu}_i & = &  - G  \sum_{j \in \inn/\out }     \frac{\big(2 (p_i \cdot p_j)^3-3 m_i^2 m_j^2 \, p_i \cdot p_j \big)p_j^\mu -m_i^2 m^4_j p^\mu_i}{\big((p_i \cdot p_j)^2- m_i^2 m_j^2\big)^{3/2}}  \\
& = &   - G  \sum_{\substack{j \in \inn/\out \\ m_j \neq 0}}     \frac{\big(2 (p_i \cdot p_j)^3-3 m_i^2 m_j^2 \, p_i \cdot p_j \big)p_j^\mu -m_i^2 m^4_j p^\mu_i}{\big((p_i \cdot p_j)^2- m_i^2  m_j^2\big)^{3/2}} + 2 G \sum_{\substack{j \in \inn/\out \\ m_j = 0}} p^\mu_j ,
\ea
where the sum is over incoming/outgoing particles for $i \in \inn/\out$, and in the second line we separated massive and  massless contributions. The above expression holds for any value of $m_i$, but  in the massless case it simplifies to
\be \label{charmmassless}
c^{\harm \, i}_\mu=   2 G  \sum_{j \in \inn/\out }   p^\mu_j      \quad \quad  (m_i = 0). 
\ee
Importantly, the log deviation vector takes the same constant value among massless outgoing/incoming particles, and can therefore be ``gauged away" by a log translation. Doing so leads to the (future/past) log translation radiative frame \cite{gianni1}, see below.

For the trajectories  \eqref{asymtraj}, the particles' asymptotic angular momentum exhibits a logarithmic  divergence \cite{laddhasen1},
 \be \label{Jmunudiv}
J_{\mu\nu}^i    \stackrel{s \to \pm \infty}{=}   \log | s | J^{\div \, i}_{\mu \nu} +  \cdots
\ee
where
\be \label{defJdiv}
J^{\div \, i}_{\mu \nu} =   c^i_{[\mu} \, p^i_{\nu]} =c^i_\mu p^i_\nu - c^i_\nu p^i_\mu .
\ee

The coefficient of this divergence  inherits the redundancy \eqref{delLci} and we denote by 
\be
J^{\harm \, i}_{\mu \nu} = c^{\harm \, i}_{[\mu} \, p^i_{\nu]} 
\ee
the expression obtained in harmonic gauge. \\

After this preliminary, we can finally write the tail formula \cite{proofdeq4} as
\bal \label{honepm}
\ov{1}{h}^+_{\mu \nu} &= 4G \left( \sum_{ i \in \out} \frac{p^i_{(\mu}  J^{\harm \, i}_{\nu)\rho} n^{\rho}}{p_i \cdot n} +  2 G P \cdot   n    \sum_i \frac{p^i_\mu p^i_\nu}{p_i \cdot n} \right)\\
\ov{1}{h}^-_{\mu \nu} &= -4G \sum_{ i \in \inn} \frac{p^i_{(\mu}  J^{\harm \, i}_{\nu)\rho} n^{\rho}}{p_i \cdot n}
\eal
where  
\be \label{totalmomentum}
P^\mu = \sum_{i \in \out} p^\mu_i = -\sum_{i \in \inn} p^\mu_i
\ee
 is the total linear momentum.

To make contact with the matching results of \cite{gianni2}, we need to express the log deviation vectors in \eqref{honepm} in the corresponding future/past radiative  frame.\footnote{Defined by the condition that outgoing/incoming null rays have no logarithmic deviation. Using \eqref{charmmassless} and \eqref{totalmomentum} one can check that $ c^{\rad \pm}_\mu$  in \eqref{charmitocrad} satisfies this condition.} Denoting them with  a label ``$\rad \pm$", one has
\be \label{charmitocrad}
c^{\harm}_\mu   = c^{\rad \pm}_\mu  \pm 2 G P_\mu \implies J^{\harm \, i}_{\mu \nu} =  J^{\rad \pm \, i}_{\mu \nu} \pm 2 G P_{[\mu} \, p^i_{\nu]}.
\ee
Substituting  \eqref{charmitocrad} in \eqref{honepm}, one finds the tail coefficients  \eqref{honepm} can be written as
\be \label{hpm1}
\ov{1}{h}^\pm_{\mu \nu} = \pm 4 G \left(\vphantom{\sum_{\out}} \right. \sum_{ \substack{ i \in \pm  \\ m_i \neq 0}} \frac{p^i_{(\mu}  J^{\rad^\pm \, i}_{\nu)\rho} n^{\rho}}{p_i \cdot n}  + 2 G P \cdot n  \sum_{i \in \inn} \frac{p^i_\mu p^i_\nu}{p_i \cdot n}+ 2 G P_\mu P_\nu \left.  \vphantom{\sum_{\out}} \right),
\ee
where we use the shorthand notation $i \in \pm$ for indices running over outgoing/incoming particles respectively. Note that the first sum in \eqref{hpm1} is restricted to massive particles since $J^{\rad^\pm \, i}_{\mu \nu}=0$ for outgoing/incoming massless particles.\\

For later purposes, we shall refer to terms proportional to $ J^{\div}_{\mu \nu}$ as ``linear''  and those  proportional  to $G P^\mu$  as  ``non-linear'', or ``drag'' terms.  The terminology refers to whether they are linear  or quadratic in the metric perturbation, although both are of the same order in $G$, since $J^{\div}_{\mu \nu}=O(G)$. Furthermore, the splitting depends  on the choice of log frame, as seen  by comparing \eqref{hpm1} with \eqref{honepm}. Nevertheless, it will occasionally be useful to make this distinction.

\section{Partially peeling Bondi metric} \label{bondipartialpeelsec}

Spacetimes that are asymptotically flat near null infinity are conveniently described by Bondi-Sachs coordinates \cite{bondi,sachs} $(r,u,\phi^A)$,  where $r$ is  a radial distance, $u$ a time variable and $\phi^A$ coordinates on the celestial sphere.  We follow  the conventions used in \cite{gianni2}, where $r$ is positive/negative  near future/past  null infinity, and  $u$ denotes  retarded/advanced time, respectively.\footnote{In terms of  $(t,r)$ coordinates, the conventions for the (Bondi frame) outgoing/incoming coordinates are: $u\atscri{\pm}=t \mp r + \cdots $ and  $r\atscri{\pm}=\pm r+ \cdots$.  \label{defretcoordfnote}}

 Bondi gauge fixing conditions read 
\be \label{bondisachsgge}
g_{rr}=g_{rA}=0, \quad \det g_{AB} = r^4 \det q_{AB},
\ee
where $q_{AB}$ is a reference $2d$ metric, used to raised and lower $2d$ indices. We will work with complex coordinates $\phi^A=(z,\zb)$ such that
\be \label{defzzb}
 q_{zz}= q_{\zb \zb}=0,
\ee
and adopt, for the most part, a unit-sphere metric (referred to as Bondi frame)
\be
q_{z \zb} = 2 /(1+ |z|^2)^{2},
\ee
although some  computations are simplified in the flat frame, where $q_{z \zb} =1$.

 In order to accommodate for generic scattering spacetimes \cite{damour,christo}, we assume Winicour's  logarithmic  asymptotic flatness conditions \cite{winicour} (see \cite{geillerpeeling} for a detailed discussion) 
\bal \label{bondimetric}
g_{AB} & \stackrel{r \to \pm \infty}{=}  r^2 q_{AB} +r \, C_{AB}\atscri{\pm} +  p_{AB}\atscri{\pm}+ \frac{1}{4} q_{AB} C^2\atscri{\pm} +O(\ln^2 r/r)\\
g_{ur} & \stackrel{r \to \pm \infty}{=}  -1 + O(1/r^2) \\
g_{uA} & \stackrel{r \to \pm \infty}{=}   \frac{1}{2} D^B C_{AB}\atscri{\pm}+  \log |r| \frac{2 G }{3 r} \, \ov{\log r}{\N_A}\atscri{\pm}+  \frac{2 G }{3 r } \Nt_A\atscri{\pm} + O(\ln^2 r/r^2), \\
g_{uu} & \stackrel{r \to \pm \infty}{=} -R/2 + \frac{2 G  \M\atscri{\pm}}{r} + O(\ln r/r^2) 
\eal
where $R$ and $D_A$ are  the scalar curvature and covariant derivative of $q_{AB}$ respectively, and $C^2\equiv C^{AB} C_{AB}$.  At the order displayed, the  only difference with the standard Bondi-Sachs metric is the presence of the trace-free tensor ${p_{AB}}\atscri{\pm}$ and the vector ${\ov{\log r}{\N_A}}\atscri{\pm}$, which we shall  refer to as ``Winicour" or ``peeling-violating" coefficients.  Einstein equations imply\footnote{Sufficiently weak fall-offs in massless matter could lead to stress-tensor terms in the RHS of \eqref{uindep}  and \eqref{Nlogritop} \cite{nichoflan,compererobert2}. }
\ba \label{uindep}
\partial_u p_{AB}\atscri{\pm} &= &0, \\
  D^B {p_{AB}}\atscri{\pm}  &= &G \, \ov{\log r}{\N_A}\atscri{\pm}.  \label{Nlogritop}
\ea

In Appendix \ref{NPapp} we review how these coefficients lead to a mild  violation of the  peeling conditions. Stronger violations can occur within the general class of ``polyhomogenous metrics''  \cite{chrusciel} of which \eqref{bondimetric} is the simplest realization, see appendix \ref{polyapp}. 

Let us now recall the  ``standard'' coefficients in \eqref{bondimetric}.  Bondi's \emph{mass aspect} (capturing the energy that came out/in at retarded/advanced time $u$ per solid angle) is given by $\pm \M\atscri{\pm}$. The \emph{shear} $C_{AB}\atscri{\pm}$  encodes the outgoing/incoming gravitational wave profile. According to the discussion of  section  \ref{tailsec} it satisfies\footnote{In section \ref{newmatching} we incorporate the effects due to incoming memory and tail terms, which are here suppressed by the  condition  \eqref{largeushearin}.}
\ba 
C_{AB}\atscri{+}(u,\phi) & \stackrel{u \to \pm \infty}{=}  & \ov{0}{C}^\pm_{AB}(\phi)  + \frac{1}{u} \ov{1}{C}^\pm_{AB}(\phi)  + \cdots, \label{largeushearout} \\
C_{AB}\atscri{-}(u,\phi) &\stackrel{u \to \pm \infty}{=} & o(1/u) . \label{largeushearin}
\ea

The coefficients ${\Nt_A}\atscri{\pm} $ contain information on the angular momentum leaving/coming into spacetime at retarded/advanced time $u$. There are various prescriptions  for how to isolate the  \emph{angular momentum aspect}  ${\N_A}\atscri{\pm}$  \cite{compnich}. We follow the same convention as in \cite{gianni2}, where
\be \label{defNA}
 \Nt_A =  \N_A +u \partial_A \M -\frac{3}{32 G}\partial_A C^2 +\frac{u}{4 G} D^{B} D_{[B}D^{C}C_{A]C}.
\ee
Einstein equations then imply
\be \label{dotangmomasp}
\partial_u {\N_z}\atscri{\pm} = - \frac{1}{2G} u  \partial_u  D_z^3 {C^{ zz}}\atscri{\pm} +  \text{angular momentum flux},
\ee
where ``angular momentum flux'' stands for terms that are quadratic in the shear, of the type  $O(C \partial_u C /G)$ and $O(u (\partial_u C)^2/G)$, as well as possible massless matter contribution, see e.g. \cite{HPS}.  We shall only be interested in the linear term in \eqref{dotangmomasp}, as it is the one capturing the asymptotic behavior of the angular momentum aspect:  Substituting \eqref{largeushearout}, \eqref{largeushearin}  in \eqref{dotangmomasp} leads to
\ba 
\N_z\atscri{+} & \stackrel{u \to \pm  \infty}{=} &\log | u | \ov{\log u}{\N_z}^\pm   + O(u^0),  \label{largeuangaspout} \\
\N_z\atscri{-} & \stackrel{u \to \pm  \infty}{=} & O(u^0),   \label{largeuangaspin}
\ea
with
\be \label{NzloguitoCone}
\ov{\log u}{\N_z}^\pm=   \frac{1}{2 G} D_z^3 \ov{1}{C}^{\pm \, zz} .
\ee

\section{Celestial CFT detour} \label{cftsec}

The Bondi metric  \eqref{bondimetric} has an intrinsic Weyl-rescaling  freedom,\footnote{For simplicity we are regarding $\lambda$ to be a constant rescaling factor,  but it can be promoted into a local  one (in $\phi^A$ and/or $u$)  by  ``Weyl-covariantizing''   the metric coefficients  \cite{BT,bmsw,geillzwikpartial}. The resulting Weyl rescalings underlie  the conformal approach to asymptotically flat spacetimes \cite{penrosezrm,geroch}.} 
\bal   \label{scalingruphi}  
&\delta r = -\lambda r,  \quad \delta u = \lambda u, \quad \delta \phi^A = 0 ,\\
& \delta q_{AB} = 2 \lambda q_{AB},  \quad  \delta C_{AB} =\lambda(1-u \partial_u) C_{AB},  \quad \delta p_{AB} =  0, \quad \text{etc.}
\eal
under which the spacetime line element $ds^2$ is left unchanged. Denoting by  $[\O]$ the Weyl weight of a given quantity $\O$,
\be \label{defsw}
\delta \O = -\lambda [\O] \O,
\ee
we  have\footnote{Fields depending on $u$, such as the shear,  have  operator-valued Weyl weights (see e.g. \cite{2dstress,discretebasis,crestothesis}) that can be diagonalized by a Mellin transform \cite{conformalbasis}. } 
\bal  
&[r] = 1, \quad  [u] = -1, \quad [\phi^A] = 0 ,\\
& [q_{AB}] = -2,  \quad  [C_{AB}] =-1+u \partial_u,  \quad [p_{AB}] =  0.
\eal

The  freedom \eqref{scalingruphi} is  typically fixed by the choice of  $q_{AB}$ in  \eqref{bondimetric}.  The scaling properties of the different quantities  are, however, manifested in the action of the asymptotic Lorentz group,
famously acting as two-dimensional conformal transformations  \cite{sachs}.  This symmetry, together with its local extension \cite{BTprl}, lies at the core of  the celestial holography program \cite{raclariulects,pastelects,donnayreview},  where asymptotic fields are interprated as operators in a putative CFT  on the celestial sphere.   In that setting, fields are characterized by their  \emph{conformal dimension} $\Delta$, which for a  symmetric and traceless tensor $\O_{A_1, \ldots, A_s}$ is related to the Weyl weight \eqref{defsw} by 
\be
\Delta_{\O_{A_1, \ldots, A_s}} = [\O_{A_1, \ldots, A_s}] + s.
\ee

Among the Bondi fields discussed in section \ref{bondipartialpeelsec}, the ones we will be interested in are: the shear tail $\ov{1}{C}_{AB}$ \eqref{largeushearout}, the logarithmic angular aspects $\ov{\log r}{\N_A}$ \eqref{bondimetric} and $ \ov{\log u}{\N_A}$ \eqref{largeuangaspout}  and the Winicour tensor ${p_{AB}}$ \eqref{bondimetric}.   As we will see in the next section, these coefficients depend on the scattering momenta through certain elementary “soft tensors”  we now describe.

Given a  4-momentum $p^\mu$ and a 4-vector $b^\mu$ let us define
\bal \label{def4softtensors} 
s_z &: =  \log( |p \cdot n|) \partial_z n^\mu  n^\nu b_{[\mu}p_{\nu]},  \\ 
s_{zz} &:=   \frac{ p \cdot \partial_z n}{p \cdot n}  \partial_z n^\mu n^{\nu} b_{[\mu} p_{\nu]}  \\ 
\st_{zz} &:= \Big(2 \frac{(  p \cdot \partial_z n)^3}{ (p \cdot n)^3} q^{z \zb}\partial_{\zb} n^\mu  n^\nu  + 3  \frac{  p \cdot \partial_z n}{ (p \cdot n)^2}  \partial_z n^\mu  p^\nu\Big) b_{[\mu}p_{\nu]} \\
\st_z &: =   - \frac{3}{2}  \frac{m^4  }{(p \cdot n)^4}  \partial_z n^{\mu} n^{\nu} b_{[\mu} p_{\nu]} 
\eal
where $n^\mu$ is the null vector associated with the point $(z,\zb)$ on the celestial sphere and  $m^2 \equiv - p \cdot p$. The expressions are valid for both massive and massless momenta, with $\st_z$ becoming a distribution in the massless limit, see Eq. \eqref{stzmassless}. Their Weyl weights  follow  from  $[n^\mu]=-1$, $[q^{z \zb}]=2$ and coincide with those of the associated Bondi coefficients, see Table \ref{softtensors} and the end of this section for further details.

\begin{table}[h!]
\centering
\begin{tabular}{l | c | c | c}
 soft tensor  & $[ \quad ]$ & $\Delta$ & Bondi field\\
\hline
 $s_{A}$   & -2 &  -1 & ? \\ 
\hline 
 $s_{AB}$   & -2 & 0& $ \ov{1}{C}_{AB}$\\
\hline
 $\st_{AB}$    & 0 &  2 & $p_{AB}$\\
\hline
 $\st_A$  & 2 & 3 & $\ov{\log r}{\N_A}$ and $ \ov{\log u}{\N_A}$ 
\end{tabular}
\caption{Scaling properties of the soft tensors and associated  Bondi coefficients.}
\label{softtensors}
\end{table}

The vector $s_z$ is special in two respects: it does not appear to be directly associated with any Bondi coefficient and it generates all remaining soft tensors through derivatives,
\be \label{relssofttensors}
s^{zz} = D^z s^z, \quad  \st_{zz} = D^3_z s^z, \quad \st_z=   D^3_z s^{zz} = D^z \st_{zz} ,
\ee
where in the last equality we used   $[D^z,D^3_z]=0$.  The four tensors \eqref{def4softtensors}, together with the relations \eqref{relssofttensors}, form a particular instance of a \emph{celestial diamond} \cite{diamond1,diamond2}.  The associated diamond diagram is schematically given by:\footnote{The   diagram takes place in the $(J,\Delta)$ plane, with the spin $J$ (number of downstairs minus upstairs $z$ indices) increasing to the right and $\Delta$ increasing downward.}
\be \label{diamond}
\begin{tikzpicture}[baseline={(current bounding box.center)}]
  \def\L{1.3}  
  \def\l{0.9}  

  \node (A) at (0,\L) {$s^z$};
  \node (B) at (2*\L, -\L) {$\st_{zz}$};
  \node (C) at (\L,-2*\L) {$\st_z$};
  \node (D) at (-\L,0) {$s^{zz}$};

\node at (-\l,\l) {$D^z$};
\node at (-\l+2*\L,\l-2*\L) {$D^z$};
\node at (\l+\L/2,\l-\L/2) {$D^3_z$};
\node at (\l-\L/2,\l-3*\L/2) {$D^3_z$};

\draw[->] (A) --  (B);
  \draw[->] (B) --  (C);
  \draw[->] (A) --  (D);
  \draw[->] (D) --  (C);

\end{tikzpicture} \quad ,
\ee
with tensors on opposite corners related by a shadow transform \cite{kapecmitra,donnaypuhmstrom,donnaypastepuhm}.

By formally replacing $b_\mu \to i \partial/\partial p^\mu$ in \eqref{def4softtensors}, one recovers a well-studied celestial diamond relevant for tree-level quantum amplitudes, with $s^{zz}$ encoding  (the orbital contribution to)  the subleading soft graviton insertion \cite{stromingercachazo},  $\st_{zz}$  the celestial stress tensor \cite{2dstress} and $s^z$ the associated reparametrization mode \cite{repmode,srmode,pastebaner}. The vector $\st_z$ can be interpreted either as an angular momentum aspect \cite{stromvirasoro} or as a boundary-to-bulk Green’s function for superrotations \cite{clmassive}. 
 An analogous diamond describes  the 1-loop corrections to these quantities \cite{donnaynguyenruzzi,berndaviesnohle,sahoosen,1loop2dstress,agrawaldonnaynguyenruzzi,hehuangwen,pasterskiloopcomment,kadhe}. For the purposes of the present paper, we will take $b^\mu$ to be either a log deviation vector $c^\mu$ or the total linear momentum $P^\mu$.

The CFT perspective applies most naturally for massless $p^\mu$. In this case $\st_z$  becomes a distribution on the sphere that  produces the contact terms  in the   celestial stress tensor  Ward identities, see e.g. \cite{kapecmitra}. To understand this limit, it is  convenient to  rewrite  $\st_z$ as
\be \label{stzrew}
 \st_z =  - \frac{1}{2} \left(n \cdot b \partial_z + 3 \partial_z n \cdot b\right) \frac{m^4}{(p \cdot n)^3},
\ee
where the factor $m^4/(p \cdot n)^3$ is  (minus) the Bondi mass aspect associated to a particle with momentum $p^\mu$ and the action of the $b$-dependent factor can  be understood as generating an angular momentum aspect from a  translation $b^\mu$  \cite{oli,coad}.  
In the massless limit $m^4/(p \cdot n)^3 \to - 4 \pi E \delta^{(2)}(\ph,\nh)$ and  \eqref{stzrew} becomes
\be \label{stzmassless}
 \st_z \stackrel{m^2 \to 0}{=} 2 \pi E \left(n \cdot b \partial_z + 3 \partial_z n \cdot b\right) \delta^{(2)}(\ph,\nh),
\ee
where $E$ is the energy and $\ph$ the spatial direction of the massless momentum.

\subsection*{Weyl rescalings}

Let us clarify a subtlety that we have  glossed over. The first three tensors in \eqref{def4softtensors} are actually defined modulo terms in the kernels of the differential operators acting on them. As shown below, in all cases the ambiguity is proportional to $b_{[\mu}p_{\nu]}$, with a $p$-independent coefficient. The expressions for the Bondi coefficients to be discussed in the next section involve sums over incoming/outgoing particles such that
\be \label{sumbpzero}
\sum_i b^i_{[\mu}p^i_{\nu]} = 0,
\ee
and are therefore unaffected by these ambiguities. The situation is  analogous to the way angular momentum conservation ensures  gauge invariance of the subleading soft graviton factor \cite{stromingercachazo}.

The ambiguity in $s_z$ is already apparent from the logarithm in its definition \eqref{def4softtensors}, while that of $s_{zz}$ and $\st_{zz}$ can be seen by considering their behavior under local Weyl rescalings
\be \label{dellamnu}
\delta_\lambda n^\mu = \lambda(z,\zb) n^\mu.
\ee
Evaluating \eqref{dellamnu} in  \eqref{def4softtensors} leads to
\bal  \label{dellamss}
\delta_\lambda s_z & =  2 \lambda s_z    + \lambda \,\partial_z n^\mu  n^\nu b_{[\mu}p_{\nu]}, \\ 
\delta_\lambda s_{zz} &=    2 \lambda s_{zz}   +\partial_z \lambda \,  \partial_z n^\mu  n^\nu b_{[\mu}p_{\nu]} ,  \\ 
\delta_\lambda  \st_{zz}  &=   6 \partial_z \lambda  \,  \partial_z n^\mu  k^\nu b_{[\mu}p_{\nu]} ,    \\
\delta_\lambda \st_z & =   -2 \lambda \st_z ,
\eal
where, for simplicity, we computed $\delta_\lambda  \st_{zz}$ in a flat frame in which\footnote{To make sure that $\delta_\lambda  \st_{zz} \propto  b_{[\mu}p_{\nu]} $ holds in \emph{any} frame (and not just the flat one) we can evaluate the change of $\st_{zz}$ under a \emph{finite} Weyl rescaling $n^\mu \to e^{\lambda(z,\zb)} n^\mu$. One finds $\st_{zz} \to \st_{zz} + \Delta_\lambda  \st_{zz} $ with
\be \label{eqDeltastzz}
\Delta_\lambda  \st_{zz} = \left[ -2 (\partial_z \lambda)^3 \partial^z n^\mu n^\nu    +3 (\partial_z \lambda)^2 ( \partial^z n^\mu \partial_z n^\nu -n^\mu k^\nu )   +6 \partial_z \lambda  \,  \partial_z n^\mu  k^\nu \right]  b_{[\mu}p_{\nu]},
\ee
which again vanishes  under  \eqref{sumbpzero}. Since any frame can be reached from the flat one by a finite Weyl-rescaling, this establishes the desired result.
}
\be
n^\mu = \frac{1}{\sqrt{2}}\left( 1+ |z|^2, z+ \zb, -i (z- \zb), 1- |z|^2 \right),  \quad k^\mu= \partial_z \partial^z n^\mu.
\ee

The  inhomogeneous terms  in  \eqref{dellamss} capture the  ambiguities in the definition of the tensors \eqref{def4softtensors}. As anticipated, they are cancelled  by the action of the derivative operators in \eqref{diamond}, and, importantly, they are   proportional to $b_{[\mu}p_{\nu]}$ and hence vanish under \eqref{sumbpzero}.

\section{Peeling-violating  coefficients} \label{peelcoefsec} 
In \cite{gianni2} we derived,  based on the general asymptotic framework of \cite{cgw}, matching conditions for the logarithimic angular momentum aspects at   timelike and spatial infinities. The conditions  involve the \emph{sum} of $\ov{\log r}{\N_A}$ and $ \ov{\log u}{\N_A}$ at the corresponding boundaries.  In the notation of the present paper and for the case where there is no incoming soft radiation  \eqref{largeushearin}, the matching conditions are (see Eqs. (5.17) and (5.19) of \cite{gianni2})
\bal \label{ourmatching}
\ov{\log u}{\N_z}^+  + \ov{\log r}{\N}_{z}\atscri{+}  & =    - 3 \sum_{\substack{i \in \out \\ m_i \neq 0}} m_i^4 \frac{  \partial_z n^{\mu} n^{\nu} J^{\rad + i}_{\mu \nu} }{(p_i \cdot n)^4}, \\
\ov{\log r}{\N}_{z}\atscri{-} & =    - 3 \sum_{\substack{i \in \inn \\ m_i \neq 0}} m_i^4 \frac{  \partial_z n^{\mu} n^{\nu} J^{\rad - i}_{\mu \nu} }{(p_i \cdot n)^4},\\
\ov{\log u}{\N_z}^-  + \ov{\log r}{\N}_{z}\atscri{+} & =  -    \ov{\log r}{\N}_{z}\atscri{-}  + 4 G  (  n \cdot P \partial_z + 3  \partial_z n \cdot P )  \M|_{\partial_+ \I^-} ,
\eal
where 
\be \label{bondiMnoincomrad}
\M|_{\partial_+\scri^-}= -\sum_{i \in \inn}  \frac{m_i^4}{(p_i \cdot n)^3} ,
\ee
  is (minus) the  Bondi mass aspect at spatial infinity.\footnote{Eq.  \eqref{bondiMnoincomrad} only holds under \eqref{largeushearin}, otherwise there are incoming shear contributions; see  Eq. \eqref{bondiMspisoft}.}

These conditions can be   expressed in terms of the soft vector $\st_z$ as
\ba
\ov{\log u}{\N_z}^+  + \ov{\log r}{\N}_{z}\atscri{+}  & = &   2   \sum_{ \substack{ i \in \out  \\ m_i \neq 0}}  \st_{z}[p_i, c^{\rad +}_i] , \label{matchingfuture} \\
\ov{\log r}{\N}_{z}\atscri{-} & = &    2   \sum_{ \substack{ i \in \inn  \\ m_i \neq 0}}  \st_{z}[p_i, c^{\rad -}_i], \label{matchingpast}\\
\ov{\log u}{\N_z}^-  + \ov{\log r}{\N}_{z}\atscri{+} & = & -    \ov{\log r}{\N}_{z}\atscri{-}  + 8 G \sum_{i \in \inn}  \st_{z}[p_i, P], \label{matchingspi}
\ea
where the notation $\st_{z}[p,b]$ specifies the 4-vectors $p^\mu$ and $b^\mu$ in \eqref{def4softtensors}. For the last term in \eqref{matchingspi} we used \eqref{bondiMnoincomrad} and the identity \eqref{stzrew} to write
\be \label{bmassapitostz}
 (  n \cdot P \partial_z + 3  \partial_z n \cdot P )  \M|_{\partial_+ \I^-} = 2  \sum_{i \in \inn}  \st_{z}[p_i, P].
\ee

Let us now consider the tail formula \eqref{hpm1} pulled-back to the celestial sphere,  $\ov{1}{C}^\pm_{zz}   \equiv   \partial_z n^\mu \partial_z n^\nu \ov{1}{h}^\pm_{\mu \nu}$. Using again the notation from the previous section, we have
\be \label{sheartailformula}
\ov{1}{C}^\pm_{zz}  = \pm 4 G  \sum_{ \substack{ i \in \pm  \\ m_i \neq 0}}  s_{zz}[p_i, c^{\rad \pm}_i] \mp 8 G^2  \sum_{i \in \inn}  s_{zz}[p_i, P].
\ee

Combining \eqref{sheartailformula}   with  Eq.  \eqref{NzloguitoCone} leads to 
\be \label{tailsangasp}
\ov{\log u}{\N_z}^\pm= \pm 2   \sum_{ \substack{ i \in \pm  \\ m_i \neq 0}}  \st_{z}[p_i, c^{\rad \pm}_i] \mp 4 G  \sum_{i \in \inn}  \st_{z}[p_i, P].
\ee

Eq. \eqref{tailsangasp} may be thought of as a rewriting of the LSSS tail formula \eqref{hpm1}. We shall now discuss its  interplay with the matching conditions \eqref{ourmatching}.  The first observation is that the latter correctly capture the \emph{difference} of the coefficients \eqref{tailsangasp}, as can be checked by taking the linear combination $\eqref{matchingfuture}+ \eqref{matchingpast}- \eqref{matchingspi}$. This is, essentially, the  ``asymptotic derivation'' of the logarithmic soft theorem given in \cite{gianni2}.   The matching conditions are however agnostic to individual  values of $\ov{\log u}{\N_z}^\pm$, as these appear added to the unknown coefficient  $\ov{\log r}{\N}_{z}\atscri{+}$.
  Bringing in the tail formula \eqref{tailsangasp} allows one to solve for the latter in \eqref{matchingfuture} (or equivalently in \eqref{matchingspi}). The result is quite remarkable: the outgoing contributions cancel out, and one is left with the drag-type term, 
 \be  \label{Nlogrscrip}
 \ov{\log r}{\N}_{z}\big|_{\scri^+}   =   4 G \sum_{i \in \inn}  \st_{z}[p_i, P].
\ee

Eqs. \eqref{matchingpast} and \eqref{Nlogrscrip} provide  the first set of peeling-violating coefficient formulas.  Using the results from the previous section, we can  invert  Eq. \eqref{Nlogritop} to obtain the second set 
\ba 
p_{zz}\atscri{-} & =&  2 G \sum_{ \substack{ i \in \inn  \\ m_i \neq 0}}  \st_{zz}[p_i, c^{\rad -}_i]  ,  \label{pzzformulamin} \\
p_{zz}\atscri{+} & =& 4 G^2 \sum_{i \in \inn}  \st_{zz}[p_i, P] .   \label{pzzformulaplus}
\ea

As a consistency check, we now verify that \eqref{pzzformulamin} and \eqref{pzzformulaplus} reduce to  Damour's result \cite{damour} in the Newtonian limit.

\subsection{Newtonian limit}
The Newtonian limit is most naturally stated in Bondi frame, for which the null vector $n^\mu$ is
\be
n^\mu=(1,\nh).
\ee
We assume the incoming particle 4-momenta has the form
\be \label{pinewt}
p_i \approx -(m_i, \vec{\p}_i), \quad \p_i \ll m_i , \quad (i \in \inn),
\ee
where the minus sign is due to our conventions. In this limit, the dominant piece in  $\st_{zz}$ is given by the second term in  \eqref{def4softtensors},
\ba
\st_{zz} & \approx & -3 \, \frac{  \vec{\p} \cdot \partial_z \nh}{m^2} \, \partial_z n^\mu  p^\nu b_{[\mu}p_{\nu]} ,\\
& \approx & 3 \, \frac{  \vec{\p} \cdot \partial_z \nh}{m^2} (m^2 \vec{b} \cdot \partial_z \nh - \vec{\p} \cdot \partial_z \nh \, m  \, b^0), \label{stzznewt}
\ea
where  
the dot  stands for 3-vector inner product. We next need to specify  $b^\mu$. For  $p_{zz}\atscri{-}$ we take
\be
b^\mu = c^{\rad - \, \mu}_i \approx (0, \vec{c}_i)
\ee
where $\vec{c}_i$ is the Newtonian log deviation vector (see appendix \ref{damourapp}). In this case, only the first term in \eqref{stzznewt} contributes and \eqref{pzzformulamin} becomes
\be \label{newtlimpzzmin}
p_{zz}\atscri{-} \approx  6 G \sum_{i \in \inn}   \vec{\p}_i \cdot \partial_z \nh  \, \vec{c}_i \cdot \partial_z \nh.
\ee

For $p_{zz}\atscri{+}$ we  take
\be
b^\mu = P^\mu \approx (M, \vec{0}).
\ee
The relevant piece is now the second term in \eqref{stzznewt}, from which we find
\be \label{newtlimpzzplus}
p_{zz}\atscri{+} \approx -12 G^2 M \sum_{i \in \inn}  \frac{( \vec{\p}_i \cdot \partial_z \nh)^2}{m_i} .
\ee
Eqs. \eqref{newtlimpzzmin} and \eqref{newtlimpzzplus} are precisely the values of $p_{zz}\atscri{\pm}$ predicted by Damour's result, see appendix \ref{damourapp} for details.\footnote{A related Newtonian expression for \emph{future} peeling violation was proposed by Christodoulou in \cite{christo}. However, as emphasized in \cite{radu}, the formula in \cite{christo} is proportional to Damour's \emph{past} peeling violation \eqref{newtlimpzzmin}. The result obtained in \cite{radu} for $\Psi_0\atscri{+}$ is in agreement with Christodoulou's and hence in tension with  Damour's and ours.}

\section{Incoming soft radiation and a matching conjecture} \label{newmatching}

The previous  discussion left open  two questions:

\begin{enumerate}
\item Why does the future peeling violation \eqref{Nlogrscrip} only depend on incoming data?  

\item What happens if we relax \eqref{largeushearin} and allow for  incoming soft radiation?
\end{enumerate}

We now offer an answer to these questions by  reconsidering the matching conditions of \cite{gianni2}.  In the general case where there is incoming soft radiation, these conditions are\footnote{From now, on we switch notation and use $\partial_\pm \scri^+$  and  $\partial_\pm \scri^-$ to label $u \to \pm \infty$  coefficients at $\scri^+$ and $\scri^-$, respectively.}
\ba \label{futurematch}
\ov{\log r}{\N}_{z}\big|_{ \scri^+} + \ov{\log u}{\N}_{z}\big|_{\partial_+ \scri^+}  &= &2   \sum_{ \substack{ i \in \out  \\ m_i \neq 0}}  \st_{z}[p_i, c^{\rad +}_i] ,\\
\ov{\log r}{\N}_{z}\big|_{ \scri^-}  + \ov{\log u}{\N}_{z}\big|_{\partial_- \scri^-}  &=&    2   \sum_{ \substack{ i \in \inn  \\ m_i \neq 0}}  \st_{z}[p_i, c^{\rad -}_i],  \label{pastmatch} 
\ea
at timelike infinity, and
\bal
  \ov{\log r}{\N}_{z}\big|_{ \scri^+} +   \ov{\log u}{\N}_{z}\big|_{\partial_- \scri^+}  =  -   &  \ov{\log r}{\N}_{z}\big|_{ \scri^-} -    \ov{\log u}{\N}_{z}\big|_{\partial_+ \scri^-}  \\ & +  4 G  (  n \cdot P \partial_z + 3  \partial_z n \cdot P )  \M|_{\partial_+ \I^-}  \label{matchingspisoft}
\eal
at spatial infinity, 
where\footnote{The matching condition of the Bondi mass aspect \cite{stromgravscatt} 
\be \label{matchingbondim}
\M|_{\partial_- \scri^+} = -\M|_{\partial_+ \scri^-}
\ee
allows one to  alternatively write \eqref{bondiMspisoft}  in terms of outgoing data. The equivalence of the two expressions is a manifestation of the memory formula \eqref{diffh0} \cite{stromST}. As in \cite{gianni2} we assume a vanishing magnetic memory in the incoming soft radiation, $D_z^2   \ov{0}{C}^{zz}\big|_{\partial_\pm \scri^-}=D_{\zb}^2   \ov{0}{C}^{\zb\zb}\big|_{\partial_\pm \scri^-}$. \label{fnotemagshear}} 
\be \label{bondiMspisoft}
\M|_{\partial_+ \scri^-}  =  -\sum_{i \in \inn}  \frac{m_i^4}{(p_i \cdot n)^3}  +  \frac{1}{2 G} D_z^2  \big( \ov{0}{C}^{zz}\big|_{ \partial_+\scri^-}   -  \ov{0}{C}^{zz}\big|_{ \partial_-\scri^-}\big).
\ee

 Notice these    reduce to  \eqref{matchingfuture}, \eqref{matchingpast} and \eqref{matchingspi} upon taking  $\ov{\log u}{\N}_{z}\big|_{\partial_\pm \scri^+}  = \ov{\log u}{\N_z}^\pm$,  $\ov{\log u}{\N}_{z}\big|_{\partial_\pm \scri^-}=0$ and $\ov{0}{C}^{zz}\big|_{ \partial_\pm \scri^-}  =0$. \\

The first question is naturally answered if we postulate   \emph{two}  independent matching conditions at spatial infinity,  
\ba
  \ov{\log r}{\N}_{z}\big|_{ \scri^+}  &=& -    \ov{\log u}{\N}_{z}\big|_{\partial_+ \scri^-} + 2 G  (  n \cdot P \partial_z + 3  \partial_z n \cdot P )  \M|_{\partial_+ \I^-}   \label{spimatch1}  \\
   \ov{\log r}{\N}_{z}\big|_{ \scri^-}  &= &-  \ov{\log u}{\N}_{z}\big|_{\partial_- \scri^+} + 2 G  (  n \cdot P \partial_z + 3  \partial_z n \cdot P )  \M|_{\partial_+ \I^-}, \label{spimatch2} 
\ea
leading to   \eqref{matchingspisoft} upon addition. In the case of no incoming soft radiation, Eq. \eqref{spimatch1}  reduces to  \eqref{Nlogrscrip}.  On the other hand,  substituting \eqref{spimatch1} in  \eqref{futurematch} and \eqref{spimatch2}  in  \eqref{pastmatch}   leads to
\be \label{tailsangaspwsoftin}
\ov{\log u}{\N}_{z}\big|_{\partial_\pm \scri^+}  =  \pm 2   \sum_{ \substack{ i \in \pm  \\ m_i \neq 0}}  \st_{z}[p_i, c^{\rad \pm}_i]    \mp 2 G  (  n \cdot P \partial_z + 3  \partial_z n \cdot P )  \M|_{\partial_+ \I^-} +   \ov{\log u}{\N}_{z}\big|_{\partial_\pm ,\scri^-}   ,
\ee
which, in the absence of incoming soft radiation, reduces to the tail formula \eqref{tailsangasp} upon using \eqref{bmassapitostz}.

Thus,  the answer to  the first question (a new matching at spatial infinity)  naturally answers the second question. Namely,  \eqref{pastmatch}, \eqref{spimatch1} and \eqref{tailsangaspwsoftin} provide the modification to the formulas \eqref{matchingpast},  \eqref{Nlogrscrip} and \eqref{tailsangasp} due to incoming soft radiation.   Such type of corrections where discussed in \cite{gianni2} in the context of the logarithmic soft theorem. By the same type of argument given there, one can show that \eqref{tailsangaspwsoftin} implies the LSSS tail formula \eqref{honepm} is corrected by 
\be \label{correctedtailincomingsoft}
  \ov{1}{h}^\pm_{\mu \nu} \to \ov{1}{h}^\pm_{\mu \nu} +   \ov{1}{h}_{\mu \nu}\big|_{\partial_\pm \scri^-} \mp 2  G \, n \cdot P \, \big( \ov{0}{h}_{\mu \nu}\big|_{\partial_+ \scri^-}  - \ov{0}{h}_{\mu \nu}\big|_{\partial_- \scri^-}  \big)      ,
\ee
where  $\ov{0}{h}_{\mu \nu}\big|_{\partial_\pm \scri^-}$ and $\ov{1}{h}_{\mu \nu}\big|_{\partial_\pm \scri^-}$ are the memory and tail coefficients of the incoming waveform.  From a perturbative perspective, the first correction in \eqref{correctedtailincomingsoft} comes from the linear propagation of the incoming soft radiation, while the second one accounts for the drag  caused by the total spacetime momentum.

The proposed matching conditions  \eqref{spimatch1} and \eqref{spimatch2} also admit a simple perturbative interpretation: The first terms in the RHS are precisely what is expected in the linearized theory (see  \cite{sayalimatching,fontainematching} for the analogous matching in the case of the electromagnetic field) while the second terms represent a drag or non-linear correction. 

The distinction between these two kind of contributions to the peeling breakdown was already given by Damour in \cite{damour} (see also \cite{kehrberger4}) who emphasized the difference between  past peeling violation (occurring at linearized level) and future peeling violation (given by a non-linear effect in the absence of incoming soft radiation). 

\subsection{Comparison with Comp\`ere-Robert conditions}

In \cite{compererobert2}, Comp\`ere and Robert (CR) presented new matching conditions at spatial infinity, whose relation to our proposal  we now discuss. Their conditions take the form of  \emph{conservation laws} in the sense of identifying a quantity whose value at future and past null infinity match, as in the case of the Bondi mass aspect \eqref{matchingbondim}.  To facilitate comparison, let us start by recasting our  conditions in this form. Motivated by the  CR analysis, let us define\footnote{The $\pm$ sign in the second term of \eqref{defDz} disappears once the total momentum is expressed in terms of the future/past Bondi mass aspect,  $P^\mu = \pm \tfrac{1}{4 \pi} \int_{S^2} \, n^\mu    \M|_{\partial_\mp \I^\pm} $.}
\be \label{defDz}
\D_z\atscri{\pm} :=  \ov{\log r}{\N}_{z}\big|_{ \scri^\pm} + \ov{\log u}{\N}_{z}\big|_{\partial_\mp \scri^\pm} \pm 2 G  (  n \cdot P \partial_z + 3  \partial_z n \cdot P )  \M|_{\partial_\mp \I^\pm}  .
\ee
In terms of these quantities, the matching condition \eqref{matchingspisoft} can be written as
\be \label{matchingDz}
\D_z\atscri{+} =  - \D_z\atscri{-}.
\ee

A second independent condition can be obtained by taking the difference between \eqref{spimatch1} and \eqref{spimatch2}, leading to
\be \label{2ndmatching}
 \ov{\log u}{\N}_{z}\big|_{\partial_- \scri^+}  -  \ov{\log r}{\N}_{z}\big|_{ \scri^+}   =      \ov{\log u}{\N}_{z}\big|_{\partial_+ \scri^-} - \ov{\log r}{\N}_{z}\big|_{ \scri^-} .
\ee

Eqs.  \eqref{matchingDz} and \eqref{2ndmatching} provide a ``conservation law''  rewriting of our proposed matchings   \eqref{spimatch1} and \eqref{spimatch2}.  The CR conditions, on the other hand, can be written in our conventions as\footnote{We are omitting their matching on the dual mass aspect. In our case it becomes $0=0$ due to the purely electric boundary condition  on the  shear; see footnote \ref{fnotemagshear}.}
\be
\begin{array}{ccc}
\D_z\atscri{+} &=&  - \D_z\atscri{-} \\
& &\\
 \ov{\log u}{\N}_{z}\big|_{\partial_- \scri^+}    &= &     \ov{\log u}{\N}_{z}\big|_{\partial_+ \scri^-} 
\end{array} \quad \quad \quad \text{(CR conditions)}
\ee

Their first matching precisely coincides with ours \eqref{matchingDz}.  The second, however, differs due to the absence of $\log r$ angular  aspects and is therefore in tension with the  LSSS tail formula, as  acknowledged in \cite{compererobert2}.  We emphasize that  \eqref{2ndmatching} may be thought of as a rewriting of the  tail formula (once the other matching conditions are taken into account), but that we  currently lack an independent asymptotic derivation of this relation. It would  be important to revisit the asymptotic analysis of \cite{compererobert2} to pinpoint the origin of this tension.

\section{A comment on quantum terms} \label{quantumsec}

In perturbative quantum gravity, the shear field $C_{AB}$ can be written as  \cite{aabook,stromST}
\be \label{quantumCzz}
\Chat_{zz}(u,\phi)  = \frac{\kappa}{8 i \pi^{2}} \int_{0}^{\infty} d \w \, \left( a_{+}(\w,\phi) e^{-i\omega u} - a_{-}^{\dagger}(\w,\phi) e^{i\omega u} \right),
\ee
where   $a_\pm(\w,\phi)$ are the  (positive/negative helicity)  graviton Fock operator  with  momentum $p^\mu = \w \, n^\mu(\phi)$; $\kappa= \sqrt{32 \pi G}$ and  for simplicity we assume a flat 2d frame.  In the context of quantum electrodynamics, it was noted in  \cite{logwalok}   that operators of the type \eqref{quantumCzz} develop a logarithmic ``quantum tail'' at large $u$,\footnote{This tail can alternatively   be understood   as arising from Feynman  (as opposed to retarded or advanced) boundary conditions \cite{sayalifeynman,KrausQED,KrausYM}.} 
\be
\Chat_{zz}(u,\phi) \stackrel{u \to \pm \infty} = \log |u| \B_{zz}(\phi) + \cdots  
\ee
with
\be \label{defB}
 \B_{zz}(\phi) =  \frac{i\kappa}{4\pi^{2}} \lim_{\omega\to 0} \omega\left( a_{+}(\omega,\phi) - a_{-}^{\dagger}(\omega,\phi) \right).
\ee

In gravity, the operator \eqref{defB} was introduced and studied in \cite{bissidonnayvalsesia}, from which we  borrow notation. In spite of its formal character,  the operator yields a finite value when inserted in  the $S$ matrix, thanks to Weinberg's pole \cite{stromST}.

If we now consider the quantum metric fluctuation at linear order, it is straightforward to see that a $\log |u|$ term in the shear at past null infinity, leads to a $\log r$ term in the shear at future null infinity  (and vice versa) \cite{briceno}. We therefore conclude that
\be \label{ghat}
\ghat_{zz} \approx r \log r \,  \B_{zz} +  r \, \Chat_{zz} + \cdots.
\ee

As argued in appendix  \ref{polyapp}, a theorem due to Valiente Kroon \cite{kroon1,kroon2} implies that \eqref{ghat} leads to a ``strong'' peeling violation
\be
\Psi_0=O(r^{-3}),  \quad  \Psi_1=O(r^{-3}), \quad \Psi_2 = O(  r^{-3}  \log r).
\ee

The resulting decay in $\Psi_1$ coincides with the strong peeling violation found in \cite{radu}. The perturbative order at which their effect occurs  also match with what is expected from \eqref{ghat}: When evaluating the insertion of the operator \eqref{defB} in the $S$-matrix, one gets an extra power of $\kappa$ from Weinberg’s soft graviton theorem, resulting in an extra power of $G$ relative to the order at which the $S$-matrix is being evaluated. In particular, starting from the tree-level waveform, one obtains a one-loop corrected waveform, in agreement with the order of their effect.\footnote{When regarding  initial and final momenta as independent quantities,  tree-level and 1-loop waveforms are of order $O(G)$ and $O(G^2)$ respectively. In \cite{radu}, the final momenta  are written as a $O(G)$ deviation from the initial ones, resulting in $O(G^2)$  and $O(G^3)$ tree-level and 1-loop waveforms, respectively.}


Let us conclude by noting that the analogue of $\B_{zz}$ in electrodynamics leads to the quantum part of the log soft photon theorem \cite{sahoosen} in an ``asymptotic'' derivation of the latter \cite{logwalok}. Whereas no such derivation has yet been given in the gravitational case, we expect  a similar mechanism to be at play there. The absence of explicit  $\hbar$-factors in these quantum terms \cite{sahoosen} makes their classical limit a subtle issue \cite{manughoshladdhaathira}, and may explain their persistence in the computation of \cite{radu}.

\section{Discussion and outlook} \label{discussionsec}
In this paper we studied the  (leading) tail  and  peeling-violating coefficients of the Bondi metric at null infinity,  as captured by the $2d$ tensors
\be
\begin{array}{cc}
\text{Leading tail coefficients at $\scri^\pm$:} &  \quad  \ov{1}{C}_{zz}\big|_{\partial_\pm \I^+}, \quad    \ov{1}{C}_{zz}\big|_{\partial_\pm \I^-}   \\
& \\
 \text{Peeling-violating coefficients at $\scri^\pm$:} &   \quad p_{zz}\atscri{\pm} .
\end{array}
\ee

 Rather than dealing directly with these coefficients, our approach relied on working   with their associated \emph{logarithmic  angular momentum aspects},
\be \label{logaspects}
\ov{\log u}{\N_z}\big|_{\partial_\pm \I^+}, \quad    \ov{\log u}{\N_z}\big|_{\partial_\pm \I^-} , \quad \ov{\log r}{\N_z}\atscri{\pm}  ,
\ee
defined by
\be \label{logaspectsitoCandp}
\ov{\log u}{\N_z} =  \frac{1}{2 G} D_z^3 \ov{1}{C}^{ zz} , \quad \quad  \ov{\log r}{\N_z}   =  \frac{1}{G} D^z {p_{zz}},
\ee
at the corresponding boundaries.

In previous work \cite{gianni2}, we found  three independent relations among the  quantities \eqref{logaspects}, arising from Einstein equations near future/past timelike and spatial infinities.\footnote{The relation at spatial infinity was independently found in \cite{compererobert2}.}
Here  we showed that the gravitational  wave tail formula of \cite{laddhasen1,laddhasen,sahoosen,proofdeq4} implies a second condition at spatial infinity, leading to a total of four constraints in \eqref{logaspects}. These  can be used to determine the tail  $\ov{1}{C}_{zz}\big|_{\partial_\pm \I^+}$  and peeling violating coefficients $p_{zz}\atscri{\pm}$ in terms of scattering data.\footnote{Here we are regarding $ \ov{1}{C}_{zz}\big|_{\partial_\pm \I^-} $ and  $ \ov{0}{C}_{zz}\big|_{\partial_\pm \I^-} $  as part of the scattering data, although usually  these are  taken to be vanishing.} By construction, the resulting tail formulas are those of \cite{laddhasen1,laddhasen,sahoosen,proofdeq4}, slightly generalized to account for incoming soft radiation. The resulting peeling-violating formulas are new and reduce, in the Newtonian limit, to a known perturbative result \cite{damour}.

Our analysis leaves many open questions for future research. Perhaps the most pressing one is a discrepancy  between our proposed second matching at spatial infinity and a similar second matching found in \cite{compererobert2}. 
A somewhat related question is the appropriate ``conservation law'' interpretation of these relations. While one of  the matchings at spatial infinity should underlie the conservation of superrotation charges discussed in \cite{donnaynguyenruzzilog,chipum2}, there is no obvious symmetry candidate for the  second one.

Although our main focus was on classical scattering, we briefly commented on the stronger peeling violations expected in the quantum case. It would be important to strengthen the argument by revisiting the matching conditions in the presence of $\log u$ “quantum tails” in the shear (see \cite{briceno} for a related discussion), and by making explicit the link between the resulting strong peeling violation and the quantum part of the logarithmic soft graviton theorem of \cite{sahoosen}.

Finally, it would be interesting to explore the implications our results may have for celestial holography.   Whereas we just   used the celestial diamond picture of \cite{diamond1,diamond2} as a tool to invert the derivatives in \eqref{logaspectsitoCandp}, it would be fascinating if there were a more fundamental aspect to this identification.

 \acknowledgments
We thank Federico Capone, Geoffrey Compère,  Alok Laddha,  Guzmán Hernández-Chifflet and Sébastien Robert  for fruitful discussions.  We acknowledge  support from Pedeciba and from ANII grant FCE-1-2023-1-175902.

\appendix

\section{Weyl scalars and partial peeling} \label{NPapp}

Peeling properties are usually framed in the Newman-Penrose  formalism \cite{NPspincoeff}, in which  the gravitational field  is described by five complex scalars $\Psi_0, \ldots, \Psi_4$ that capture the independent 
components of the  Weyl tensor in a null tetrad $(\ell^a, \ellb^a, m^a, \mb^a)$,\footnote{In this appendix we use $a,b,\ldots$ to denote spacetime indices. $\ell^a$ and  $\ellb^a$ are outgoing and incoming null vector fields  (the latter usually denoted as $n^a$) and $m^a$,  $\mb^a$ are transverse complex null vector fields; see Eq. \eqref{tetradpm} for their explicit asymptotic form in Bondi coordinates.}

\bal  \label{defPsi}
\Psi_0 & = - W_{a b c d} \ell^a m^b \ell^c m^d  \\
\Psi_1 & = - W_{a b c d} \ell^a \ellb^b \ell^c m^d  \\
\Psi_2 & = - W_{a b c d} \ell^a m^b \mb^c \ellb^d  \\
\Psi_3 & = - W_{a b c d}  \ellb^a \ell^b \ellb^c \mb^d \\
\Psi_4 & = - W_{a b c d} \ellb^a \mb^b \ellb^c \mb^d  .
\eal

In terms of these quantities, the peeling conditions at future/past null infinity are  \cite{sachsoutgoing}
\be \label{peelingconditions}
{\Psi_k}\atscri{+} =   O(1/r^{5-k}), \quad \quad 
{\Psi_k}\atscri{-} = O(1/r^{1+k}), \quad \quad k=0, \ldots, 4,
\ee
where  future and past  Weyl scalars with the same decay are related by 
\be \label{psipm}
{\Psi_k}\atscri{+} \sim  {\Psi_{4-k}}\atscri{-},
\ee
 due to the  formal exchange of roles between outgoing and incoming null directions, see e.g. \cite{bardeenpress}.

The peeling conditions are known to be satisfied by the Bondi metric in the absence of Winicour terms \cite{sachs}. Once these are switched on, a mild deviation  occurs  in \cite{winicour,kroonout,geillerpeeling}, 
\be \label{winicourpeeling}
\begin{array}{ccc}
{\Psi_0}\atscri{+} , &  {\Psi_4}\atscri{-}  &= O(1/r^{4}) ,  \\
&&\\
{\Psi_1}\atscri{+},  &  {\Psi_3}\atscri{-} &=  O(\log |r|/r^{4}),
\end{array}
\ee
with all the remaining Weyl scalars obeying \eqref{peelingconditions}.  To see how \eqref{winicourpeeling} comes about, let us consider the asymptotic form of the null tetrad at future/past null infinity. In the conventions of section \ref{bondipartialpeelsec}  and in Bondi frame, these read\footnote{The expressions at $\scri^+$ are the standard ones, see e.g. \cite{geillerpeeling}. To get the expressions at $\scri^-$ we  write, following \cite{damour}, the null vectors in asymptotic coordinates $(t,r)$ such that  $\ell^a \partial_a =  {\partial_t} + {\partial_r} + \cdots$ and   $\ellb^a \partial_a = \frac{1}{2}\left({\partial_t} -  {\partial_r} \right)+ \cdots $, and then change to advanced coordinates, see footnote \ref{defretcoordfnote}.}.
\be \label{tetradpm}
\begin{array}{cll}
{\ell^a \partial_a}\atscri{+} &= \partial_r +  \cdots,   \quad   {\ell^a \partial_a}\atscri{-} &= 2 \partial_u + \cdots, \\
&&\\
{\ellb^a \partial_a}\atscri{+} & =\partial_u + \cdots,   \quad {\ellb^a \partial_a}\atscri{-}  &=  \frac{1}{2} \partial_r + \cdots,\\ 
&&\\
{m^a \partial_a}\atscri{\pm}  & =   \frac{1}{r \sqrt{q_{z\zb}}} \partial_{z} + \cdots.& 
\end{array}
\ee

 Substituting  \eqref{tetradpm} in \eqref{defPsi} leads to
\bal \label{psi04itoW}
{\Psi_0}\atscri{+} & = - \frac{1}{r^2 q_{z\zb}} W_{r z r z}\atscri{+} + \cdots ,  \\
{\Psi_1}\atscri{+} & = - \frac{1}{r \sqrt{q_{z\zb}}} W_{r u r z}\atscri{+} + \cdots ,  \\
{\Psi_3}\atscri{-} & = - \frac{1}{2 r \sqrt{q_{z\zb}}} W_{r u r \zb}\atscri{-}+ \cdots, \\
{\Psi_4}\atscri{-} & = - \frac{1}{4 r^2 q_{z\zb}} W_{r \zb r \zb}\atscri{-}+ \cdots . 
\eal

The peeling violation is therefore captured by the Weyl tensor components $W_{r z r z}$ and $W_{r u r z}$ (and their complex conjugates). For the line  element \eqref{bondimetric} these quantities are  given by  \cite{geillerpeeling}
\ba  \label{Wrzrz}
 W_{r z r z}\atscri{\pm}& =&- \frac{p_{zz}\big|_{\scri^\pm} }{r^2} + \cdots, \\
 && \nonumber\\
 W_{r u r z}\atscri{\pm}& =&- \frac{\log |r|}{r^3} \,\ov{\log r}{\N_z}\atscri{\pm}  + \cdots,
\ea
which, upon substitution in \eqref{psi04itoW} leads to  \eqref{winicourpeeling}, with specific $u$-independent coefficients.

\section{Tails and peeling breakdown in the Newtonian limit} \label{damourapp}

In the Newtonian limit, the spatial components of the gravitational waveform \eqref{defhout} are given by the linearized, quadrupolar field (see e.g. \cite{blanchetlrr}),
\be \label{hquadr}
h^\quadr_{i j}(u,\nh) = 2 G \partial^2_u Q_{ij}(u),
\ee
where $Q_{ij}$ is the quadrupole moment of the scattering particles,\footnote{In this appendix we omit  particle index labels to avoid confusion with spatial Cartesian indices $i,j,\ldots$. As in section \ref{tailsec}, we also leave implicit the transverse-traceless projection in \eqref{hquadr}.}
\be \label{defQij}
Q_{ij}(t) = \sum m \, x_{\langle i}(t) \, x_{j \rangle}(t).
\ee

At early and late times, the particle trajectories take the asymptotic form
\be \label{xtnewt}
x_i(t) \stackrel{t \to \pm \infty}{=} t \, v_i^\pm + \ln |t| c_i^\pm  + \cdots
\ee
where $c_i$ is the Newtonian  log deviation vector (see the end of the appendix for details).  Substituting in \eqref{defQij} we recover Eq. (4.9) of \cite{damour}
\be \label{Qijlarget}
Q_{ij}(t)\stackrel{t \to \pm \infty}{=}  \frac{t^2}{2} A^\pm_{ij} +t \ln |t| B^\pm_{ij} + \cdots
\ee
with
\bal \label{defABij}
A^\pm_{ij} &=  2 \sum m v^\pm_{\langle i} v^\pm_{j \rangle}, \\
B^\pm_{ij} &=  2 \sum m v^\pm_{\langle i} c^\pm_{j \rangle}. 
\eal
Plugging \eqref{Qijlarget} in \eqref{hquadr} leads to
\be \label{hquadrlimu}
h^\quadr_{i j}(u,\nh) \stackrel{u \to \pm \infty}{=} 2 G \big( A^\pm_{ij} + \frac{1}{u} B^\pm_{ij} + \cdots \big).
\ee

One can verify that \eqref{hquadrlimu}  correctly reproduces the  Newtonian limits of the memory \eqref{diffh0} and tail \eqref{honepm} formulas, up to the  ``drag term'' in the latter which, in the post-Newtonian setting, arises from the  ``monopole-quadrupole'' non-linear  correction to \eqref{hquadr} \cite{hereditary}.

Damour's peeling violation formulas, given in Eqs.  (4.10) and  (4.42)  of \cite{damour} are 
\bal \label{damourformulas}
{\psi_4}\big|_{\scri^-} & \approx \frac{3 }{4 r^4} G B^-_{i j} \mbb^i \mbb^j + \cdots \\
{\psi_0}\big|_{\scri^+} & \approx - \frac{6 }{r^4} G^2 M A^-_{i j} \m^i \m^j  + \cdots 
\eal
where, in our notation $\m^i$ is given by
\be
\m^i  =  \frac{1}{\sqrt{q_{z\zb}}} \partial_{z} \nh^{i} 
\ee
 and $M$ is the total mass. Using Eqs. \eqref{psi04itoW} and \eqref{Wrzrz}, these can be translated into formulas for the Winicour tensor, 
\bal
p_{zz}\big|_{\scri^-} &\approx  3 G   \partial_z n^i \partial_z n^j  B^-_{i j} ,\\
p_{zz}\big|_{\scri^+} &\approx  -6 G^2 M   \partial_z n^i \partial_z n^j  A^-_{i j} .
\eal

To make contact with the results of section \ref{peelcoefsec}, let us finally  express the incoming coefficients \eqref{defABij} in terms of  Newtonian 3-momenta $\vec{\p} = m \vec{v} $ and 3d inner products,
\bal
p_{zz}\big|_{\scri^-} &\approx   6 G \sum_{\substack{ \inn}} (\vec{\p} \cdot \partial_z \nh ) (\vec{c} \cdot \partial_z \nh)  ,\\ 
p_{zz}\big|_{\scri^+} &\approx  -12 G^2 M  \sum_{\substack{\inn}} \frac{(\vec{\p} \cdot \partial_z \nh)^2}{m} .
\eal
The resulting expressions  coincide with the Newtonian limits \eqref{newtlimpzzmin} and \eqref{newtlimpzzplus} of the general formulas \eqref{pzzformulamin} and \eqref{pzzformulaplus}.

\subsection*{Log deviation in the Newtonian limit}
We now briefly  discuss the Newtonian formula for the log deviation 3-vector  in \eqref{xtnewt} and  show  how it is recovered from  the relativistic expression \eqref{charm}. We  revert to the notation used in the main text, where $i,j$ stand for particle labels.

Newton's 2nd law for the $i$-th particle reads
\be \label{2ndlaw}
\frac{d^2}{d t^2} \vec{x}_i(t) = G \sum_j  \frac{m_j}{|\vec{x}_j(t)-\vec{x}_i(t)|^{3}} (\vec{x}_j(t)-\vec{x}_i(t))
\ee
Substituting 
\be \label{xtnewt}
\vec{x}_i(t) \stackrel{t \to \pm \infty}{=} t \, \vec{v}_i^\pm + \ln |t| \vec{c}_i^\pm  + \cdots
\ee
in  \eqref{2ndlaw} leads to
\be \label{newtc}
\vec{c}_i^\pm = \mp G \sum_j  \frac{m_j}{|\vec{v}_j-\vec{v}_i|^{3}} (\vec{v}_j-\vec{v}_i).
\ee

Consider now \eqref{charm} for 
\be \label{newtpmu}
p^\mu_i = \pm m_i ( \sqrt{1+v^2_i} , \vec{v}_i)  \approx \pm m_i ( 1+ v^2_i/2 , \vec{v}_i) , \quad i \in \out/\inn .
\ee
In the small velocity limit, the 4-vector dot product among outgoing/incoming momenta can then be written as
\be \label{newtpipj}
p_i \cdot p_j \approx - m_i m_j (1+ |\vec{v}_j-\vec{v}_i|^2/2) + \cdots.
\ee
Substituting \eqref{newtpmu} and \eqref{newtpipj} in \eqref{charm} leads to  \eqref{newtc}, to leading order in the velocity. 

We have focused on the spatial component of the log deviation vector in harmonic frame. The same result however holds  in the radiative frames, since, in the Newtonian limit, their difference is a purely temporal  shift \eqref{charmitocrad} proportional to $P^\mu \approx (M , \vec{0})$.

\section{Classification of non-peeling decays} \label{polyapp}

In \cite{kroon1,kroon2}, Valiente Kroon classifies the possible $ \log^n r/r^k$-type decays in  the Weyl scalars that are  allowed by Einstein equations and asymptotic flatness.\footnote{For definitiveness in what follows we focus on future null infinity, but entirely parallel considerations apply to past null infinity, with indices shifted according to \eqref{psipm}.} As in the classic analysis of Newman and Penrose \cite{NPspincoeff}, the decay of $\Psi_0$ fixes that of the remaining Weyl scalars. Valiente Kroon identifies  two families of peeling-violating fall-offs, distinguished by the behaviour of the shear at null infinity:
\be \label{1stfamily}
\Psi_0 = O(\log^n r/r^4) \implies \left\{
\begin{array}{ll}
 \Psi_1 &= O(\log^{n+1} r/r^4), \\
 &\\
  \Psi_k  &=   O(1/r^{5-k}) , \quad k \geq 2
  \end{array}\right. \quad \quad \text{(finite shear)}
\ee
and
\be \label{2ndfamily}
\Psi_0 = O(\log^n r/r^3) \implies \left\{
\begin{array}{ll}
 \Psi_1 &= O(\log^{n} r/r^3), \\
 &\\
 \Psi_2 &= O(\log^{n+1} r/r^3) \\
 &\\
  \Psi_k  &=   O(1/r^{5-k}) , \quad k \geq 3
  \end{array}\right. \quad \quad \text{(divergent shear)}
\ee
where $n$ can be any  natural number.   

To make contact with the rest of the paper, let us  discuss how the asymptotic form of $\Psi_0$  is determined from the metric components.  In  Bondi coordinates,  radial decays are set by the specification of $g_{AB}(r)$ at a given retarded time $u$, see e.g. \cite{kroonout,BT}. Since the leading peeling-violating terms are $u$-independent \cite{chrusciel,kroon1,kroon2,geillerpeeling}, we can consider the ansatz
\be \label{gralgAB}
g_{AB} \approx r^2 q_{AB} +r \, C_{AB}(u,\phi) + \D_{AB}(r,\phi)   + \frac{1}{4} q_{AB} C^2 + \cdots
\ee
where $\D_{AB}$ is traceless and the approximate equality means we only keep  leading terms in the large $r$ limit.  The two families then correspond to the choices
\be \label{Ccases}
\D_{AB}(r,\phi) =  \left\{
\begin{array}{ll}
 \log^{n} r  \, \ov{n}{\P}_{AB}(\phi) & \quad \text{(finite shear)} \\
 &\\
  r  \log^{n+1} r  \, \ov{n}{\Q}_{AB}(\phi) &  \quad  \text{(divergent shear)}
  \end{array}\right. .
\ee

To see  the relation between \eqref{Ccases} and $\Psi_0$,  consider the asymptotic form of the Weyl  tensor component 
\be \label{Wrzrzgral}
 W_{r z r z}  \approx  -\frac{1}{2} r \partial_r^2 (r^{-1}g_{zz}) .
 \ee
Substituting \eqref{Ccases} in \eqref{Wrzrzgral} gives
\be \label{Wrzrzcases}
 W_{r z r z}  \approx     \left\{ \begin{array}{l}
 - \frac{\log^{n} r}{r^2}  \, \ov{n}{\P}_{zz}    \\  
 \\
   \frac{n+1}{2 }   \frac{\log^{n} r}{r}  \, \ov{n}{\Q}_{zz}  
  \end{array}\right.
\ee
which, together with Eq.  \eqref{psi04itoW}, leads to  the  asymptotic decays in $\Psi_0$ given in  \eqref{1stfamily} and \eqref{2ndfamily}.

From the perspective presented in this paper, classical scattering spacetimes fall in the finite shear family (with $n=0$) while  quantum scattering spacetimes  belong to the divergent shear family.\footnote{The argument of section  \ref{quantumsec} leads to a  $n=0$ decay in the divergent shear family, but the conclusion relies on a linearized matching. At present we cannot rule out $n>0$ decays originating from non-linear matching terms.}


\begin{thebibliography}{99}

\bibitem{sachsoutgoing}
R.~K.~Sachs,
``Gravitational waves in general relativity. 6. The outgoing radiation condition,''
Proc. Roy. Soc. Lond. A \textbf{264}, 309-338 (1961)



\bibitem{bondi} 
  H.~Bondi, M.~G.~J.~van der Burg and A.~W.~K.~Metzner,
  ``Gravitational waves in general relativity. 7. Waves from axisymmetric isolated systems,''
  Proc.\ Roy.\ Soc.\ Lond.\ A {\bf 269}, 21 (1962).

\bibitem{sachs} 
  R.~K.~Sachs,
  ``Gravitational waves in general relativity. 8. Waves in asymptotically flat space-times,''
  Proc.\ Roy.\ Soc.\ Lond.\ A {\bf 270}, 103 (1962).


\bibitem{NPspincoeff}
E.~Newman and R.~Penrose,
``An Approach to gravitational radiation by a method of spin coefficients,''
J. Math. Phys. \textbf{3}, 566-578 (1962)


\bibitem{penrosezrm}
R.~Penrose,
``Zero rest mass fields including gravitation: Asymptotic behavior,''
Proc. Roy. Soc. Lond. A \textbf{284}, 159-203 (1965)



\bibitem{bardeenpress}
J.~M.~Bardeen and W.~H.~Press,
``Radiation fields in the Schwarzschild background,''
J. Math. Phys. \textbf{14}, 7-19 (1973)


\bibitem{cauchtorrence}
W. E. Couch and R. J. Torrence, "Asymptotic Behavior of Vacuum Space‐Times," 
J. Math. Phys. \textbf{13}, 69-73 (1978)


\bibitem{porrill}
J.~Porrill and J.~M.~Stewart,
``Electromagnetic and Gravitational Fields in a Schwarzschild Space-time,''
Proc. Roy. Soc. Lond. A \textbf{376}, 451-463 (1981)


\bibitem{novakgold}
S. Novak and J.N. Goldberg, ``Limiting behavior of asymptotically flat gravitational fields,'' 
Gen. Rel. Grav. \textbf{13}, 79–99 (1981)



\bibitem{winicour} 
J. Winicour, ``Logarithmic asymptotic flatness'',
Found. Phys. {\bf 15}, 605–616 (1985).



\bibitem{walkerwilll}
M.~Walker and C.~M.~Will,
``Relativistic Kepler problem. II. Asymptotic behavior of the field in the infinite past''
Phys. Rev. D \textbf{19}, 3495 (1979)
[erratum: Phys. Rev. D \textbf{20}, 3437 (1979)]


\bibitem{damour}
T. Damour, ``Analytical calculations of gravitational radiation'', in Fourth Marcel Grossmann
Meeting on General Relativity, pp. 365–392, Jan., 1986.


\bibitem{chrusciel}
P.~T.~Chrusciel, M.~A.~H.~MacCallum and D.~B.~Singleton,
``Gravitational waves in general relativity: 14. Bondi expansions and the polyhomogeneity of Scri,''
 Philos. Trans. A. Math. Phys. Eng. Sci. (1995) {\bf 350} (1692): 113–141 



\bibitem{kroon1}
J.~A.~ Valiente Kroon,
``Conserved quantities for polyhomogeneous space-times,''
Class. Quant. Grav. \textbf{15}, 2479-2491 (1998)

\bibitem{kroon2}
J.~A.~Valiente Kroon,
``Logarithmic Newman-Penrose constants for arbitrary polyhomogeneous space-times,''
Class. Quant. Grav. \textbf{16}, 1653-1665 (1999)

\bibitem{kroonout}
J.~A.~Valiente Kroon,
``A Comment on the outgoing radiation condition for the gravitational field and the peeling theorem,''
Gen. Rel. Grav. \textbf{31}, 1219-1224 (1999)


\bibitem{kroon3}
J.~A.~Valiente Kroon,
``Can one detect a nonsmooth null infinity?,''
Class. Quant. Grav. \textbf{18}, 4311-4316 (2001)


\bibitem{christo}
D. Christodoulou, ``The Global Initial Value Problem in General Relativity,'' in Ninth
Marcel Grossmann Meeting on General Relativity, pp. 44–54, Dec., 2002.


\bibitem{friedrichstationary}
H.~Friedrich,
``Radiative gravitational fields and asymptotically static or stationary initial data,''
[arXiv:gr-qc/0304003 [gr-qc]].


\bibitem{friedrich}
H.~Friedrich,
``Peeling or not peeling{\textemdash}is that the question?,''
Class. Quant. Grav. \textbf{35}, no.8, 083001 (2018)



\bibitem{kehrberger1}
L.~M.~A.~Kehrberger,
``The Case Against Smooth Null Infinity I: Heuristics and Counter-Examples,''
Annales Henri Poincare \textbf{23}, no.3, 829-921 (2022)


\bibitem{kehrberger2}
L.~M.~A.~Kehrberger,
``The case against smooth null infinity II: A logarithmically modified Price{\textquoteright}s Law,''
Adv. Theor. Math. Phys. \textbf{26}, no.10, 3633-3676 (2024)


\bibitem{kehrberger3}
L.~M.~A.~Kehrberger,
``The Case Against Smooth Null Infinity III: Early-Time Asymptotics for Higher $\ell $-Modes of Linear Waves on a Schwarzschild Background,''
Ann. PDE \textbf{8}, no.2, 12 (2022)



\bibitem{kehrberger4}
L.~Kehrberger,
``The case against smooth null infinity IV: Linearized gravity around Schwarzschild{\textemdash}an overview,''
Phil. Trans. Roy. Soc. Lond. A \textbf{382}, no.2267, 20230039 (2024)


\bibitem{gajic}
D.~Gajic and L.~M.~A.~Kehrberger,
``On the relation between asymptotic charges, the failure of peeling and late-time tails,''
Class. Quant. Grav. \textbf{39}, no.19, 195006 (2022)
[erratum: Class. Quant. Grav. \textbf{41}, no.11, 119501 (2024)]

\bibitem{geillerpeeling}
M.~Geiller, A.~Laddha and C.~Zwikel,
``Symmetries of the gravitational scattering in the absence of peeling,''
JHEP \textbf{12}, 081 (2024)

\bibitem{briceno}
M.~Brice{\~n}o, H.~A.~Gonz{\'a}lez, M.~Henneaux and A.~P{\'e}rez,
``Matching conditions at null infinity in the presence of logarithms: the role of advanced and retarded radiation,''
[arXiv:2510.21072 [hep-th]].



\bibitem{compererobert2}
G.~Comp{\`e}re and S.~Robert,
``A proof of conservation laws in gravitational scattering: tails and breaking of peeling,''
[arXiv:2603.08705 [hep-th]].

\bibitem{radu}
S.~De Angelis, A.~Herderschee, R.~Roiban and F.~Teng,
``Asymptotic Simplicity and Scattering in General Relativity from Quantum Field Theory,''
[arXiv:2511.10637 [hep-th]].






\bibitem{marajh}
J.~Marajh, G.~Taujanskas and J.~A.~Valiente Kroon,
``Controlled regularity at future null infinity from past asymptotic initial data: the wave equation,''
[arXiv:2508.04690 [gr-qc]].


\bibitem{bieri}
L.~Bieri,
``Radiation and Asymptotics for Spacetimes with Non-Isotropic Mass,''
Pure Appl. Math. Quart. \textbf{20}, no.4, 1601-1634 (2024)

\bibitem{godazgar}
M.~Godazgar and G.~Long,
``BMS charges in polyhomogeneous spacetimes,''
Phys. Rev. D \textbf{102}, no.6, 064036 (2020)


\bibitem{caponedg4}
F.~Capone,
``General null asymptotics and superrotation-compatible configuration spaces in $d\ge4$,''
JHEP \textbf{10}, 158 (2021)
[erratum: JHEP \textbf{02}, 113 (2022)]





\bibitem{schneiderkhera}
B.~Schneider and N.~Khera,
``From spatial to null infinity: Connecting initial data to peeling,''
Phys. Rev. D \textbf{112}, no.10, 104024 (2025)









\bibitem{DeWittBrehme}
B.~S.~DeWitt and R.~W.~Brehme,
``Radiation damping in a gravitational field,''
Annals Phys. \textbf{9}, 220-259 (1960)




\bibitem{bonnor}
W. B. Bonnor and M. A. Rotenberg, ``Gravitational waves from isolated sources,'' 
  Proc.\ Roy.\ Soc.\ Lond.\ A {\bf 289}, 247-274 (1966).


\bibitem{peters}
P. C. Peters,
``Perturbations in the Schwarzschild Metric", 
Phys. Rev. D \textbf{146}, 938 (1966)


\bibitem{ctnj}
W. E. Couch,  R. J. Torrence,  A. I. Janis and  E. T. Newman,
``Tail of a Gravitational Wave,''
J. Math. Phys. \textbf{9}, 484-496 (1968)


\bibitem{price}
R.~H.~Price,
``Nonspherical perturbations of relativistic gravitational collapse. 1. Scalar and gravitational perturbations,''
Phys. Rev. D \textbf{5}, 2419-2438 (1972)



\bibitem{hereditary}
L.~Blanchet and T.~Damour,
``Hereditary effects in gravitational radiation,''
Phys. Rev. D \textbf{46}, 4304-4319 (1992)



\bibitem{blanchetschaefer}
L.~Blanchet and G.~Schaefer,
``Gravitational wave tails and binary star systems,''
Class. Quant. Grav. \textbf{10}, 2699-2721 (1993)



\bibitem{trestini}
D.~Trestini and L.~Blanchet,
``Gravitational-wave tails of memory,''
Phys. Rev. D \textbf{107}, no.10, 104048 (2023)



\bibitem{blanchetlrr}
L.~Blanchet,
``Post-Newtonian theory for gravitational waves''
Living Rev. Rel. \textbf{27}, 4 (2024)


\bibitem{blanchetsathyaprl}
L.~Blanchet and B.~S.~Sathyaprakash,
``Detecting the tail effect in gravitational wave experiments,''
Phys. Rev. Lett. \textbf{74}, 1067-1070 (1995)



\bibitem{blanchetsathyacqg}
L.~Blanchet and B.~S.~Sathyaprakash,
``Signal analysis of gravitational wave tails,''
Class. Quant. Grav. \textbf{11}, 2807-2832 (1994)




\bibitem{stromgravscatt}
A.~Strominger,
``On BMS Invariance of Gravitational Scattering,''
JHEP \textbf{07}, 152 (2014)



\bibitem{stromST} 
  T.~He, V.~Lysov, P.~Mitra and A.~Strominger,
  ``BMS supertranslations and Weinberg’s soft graviton theorem,''
  JHEP {\bf 1505}, 151 (2015)



\bibitem{stromvirasoro}
D.~Kapec, V.~Lysov, S.~Pasterski and A.~Strominger,
``Semiclassical Virasoro symmetry of the quantum gravity $ \mathcal{S}$-matrix,''
JHEP \textbf{08}, 058 (2014)



 \bibitem{weinberg}
S.~Weinberg,
``Infrared photons and gravitons,''
Phys. Rev. \textbf{140}, B516-B524 (1965)


\bibitem{grossjackiw}
D.~J.~Gross and R.~Jackiw,
``Low-Energy Theorem for Graviton Scattering,''
Phys. Rev. \textbf{166}, 1287-1292 (1968)
 

\bibitem{white}
C.~D.~White,
``Factorization Properties of Soft Graviton Amplitudes,''
JHEP \textbf{05}, 060 (2011)



\bibitem{stromingercachazo}
F.~Cachazo and A.~Strominger,
``Evidence for a New Soft Graviton Theorem,''
[arXiv:1404.4091 [hep-th]].






\bibitem{stromym}
A.~Strominger,
``Asymptotic Symmetries of Yang-Mills Theory,''
JHEP \textbf{07}, 151 (2014)


\bibitem{2dkacmoody}
T.~He, P.~Mitra and A.~Strominger,
``2D Kac-Moody Symmetry of 4D Yang-Mills Theory,''
JHEP \textbf{10}, 137 (2016)


\bibitem{2dstress}
D.~Kapec, P.~Mitra, A.~M.~Raclariu and A.~Strominger,
``2D Stress Tensor for 4D Gravity,''
Phys. Rev. Lett. \textbf{119}, no.12, 121601 (2017)











  \bibitem{laddhasen1}
A.~Laddha and A.~Sen,
``Logarithmic Terms in the Soft Expansion in Four Dimensions,''
JHEP \textbf{10}, 056 (2018)


\bibitem{laddhasen}
A.~Laddha and A.~Sen,
``Observational Signature of the Logarithmic Terms in the Soft Graviton Theorem,''
Phys. Rev. D \textbf{100}, no.2, 024009 (2019)
[arXiv:1806.01872 [hep-th]].



\bibitem{sahoosen}
B.~Sahoo and A.~Sen,
``Classical and Quantum Results on Logarithmic Terms in the Soft Theorem in Four Dimensions,''
JHEP \textbf{02}, 086 (2019)



\bibitem{proofdeq4}
A.~P.~Saha, B.~Sahoo and A.~Sen,
``Proof of the classical soft graviton theorem in $D$ = 4,''
JHEP \textbf{06}, 153 (2020)




\bibitem{sahoosubsub}
B.~Sahoo,
``Classical Sub-subleading Soft Photon and Soft Graviton Theorems in Four Spacetime Dimensions,''
JHEP \textbf{12}, 070 (2020)



\bibitem{ghoshsahoo}
D.~Ghosh and B.~Sahoo,
``Spin-dependent gravitational tail memory in $D=4$,''
Phys. Rev. D \textbf{105}, no.2, 025024 (2022)



\bibitem{rewritten}
B.~Sahoo and A.~Sen,
``Classical soft graviton theorem rewritten,''
JHEP \textbf{01}, 077 (2022)


  
 \bibitem{senreview}
 A.~Sen,
``Gravitational wave tails from soft theorem: a short review,''
Class. Quant. Grav. \textbf{42}, no.14, 143002 (2025)






\bibitem{ciafcolfven}
M.~Ciafaloni, D.~Colferai and G.~Veneziano,
``Infrared features of gravitational scattering and radiation in the eikonal approach,''
Phys. Rev. D \textbf{99}, no.6, 066008 (2019)

\bibitem{addazibianchiven}
A.~Addazi, M.~Bianchi and G.~Veneziano,
``Soft gravitational radiation from ultra-relativistic collisions at sub- and sub-sub-leading order,''
JHEP \textbf{05}, 050 (2019)





\bibitem{bautiladdha}
Y.~F.~Bautista and A.~Laddha,
``Soft constraints on KMOC formalism,''
JHEP \textbf{12}, 018 (2022)


\bibitem{manughoshladdhaathira}
A.~Manu, D.~Ghosh, A.~Laddha and P.~V.~Athira,
``Soft radiation from scattering amplitudes revisited,''
JHEP \textbf{05}, 056 (2021)


\bibitem{vishwakarma}
S.~Paul and A.~Vishwakarma,
``Log Soft Constraints on KMOC Formalism,''
[arXiv:2601.00336 [hep-th]].


\bibitem{georgoheissrusso}
A.~Georgoudis, C.~Heissenberg and R.~Russo,
``An eikonal-inspired approach to the gravitational scattering waveform,''
JHEP \textbf{03}, 089 (2024)


\bibitem{alessiopdv}
F.~Alessio and P.~Di Vecchia,
``2PM waveform from loop corrected soft theorems,''
J. Phys. A \textbf{57}, no.47, 475402 (2024)


\bibitem{spectra}
F.~Alessio, P.~Di Vecchia and C.~Heissenberg,
``Logarithmic soft theorems and soft spectra,''
JHEP \textbf{11}, 124 (2024)


\bibitem{Bini:2024rsy}
D.~Bini, T.~Damour, S.~De Angelis, A.~Geralico, A.~Herderschee, R.~Roiban and F.~Teng,
``Gravitational waveforms: A tale of two formalisms,''
Phys. Rev. D \textbf{109}, no.12, 125008 (2024)



\bibitem{georgogoncheissjpm}
A.~Georgoudis, V.~Goncalves, C.~Heissenberg and J.~Parra-Martinez,
``Nonlinear Gravitational Memory in the Post-Minkowskian Expansion,''
[arXiv:2506.20733 [hep-th]].



\bibitem{alessiovddgonzorosi}
F.~Alessio, V.~Del Duca, R.~Gonzo and E.~Rosi,
``Gravitational amplitudes in the Regge limit: waveforms, shock waves and unitarity cuts,''
[arXiv:2601.21687 [hep-th]].





\bibitem{gianni2}
G.~Boschetti and M.~Campiglia,
``An asymptotic proof of the classical log soft graviton theorem'',
[arXiv:2603.09844 [gr-qc]]



\bibitem{blanchetpeeling}
L.~Blanchet,
``Radiative gravitational fields in general relativity. 2. Asymptotic behaviour at future null infinity,''
Proc. Roy. Soc. Lond. A \textbf{409}, 383-399 (1987)
doi:10.1098/rspa.1987.0022

\bibitem{diamond1}
S.~Pasterski, A.~Puhm and E.~Trevisani,
``Celestial diamonds: conformal multiplets in celestial CFT,''
JHEP \textbf{11}, 072 (2021)


\bibitem{diamond2}
S.~Pasterski, A.~Puhm and E.~Trevisani,
``Revisiting the conformally soft sector with celestial diamonds,''
JHEP \textbf{11}, 143 (2021)



\bibitem{kapecmitra}
D.~Kapec and P.~Mitra,
``A $d$-Dimensional Stress Tensor for Mink$_{d+2}$ Gravity,''
JHEP \textbf{05}, 186 (2018)



\bibitem{donnaypuhmstrom}
L.~Donnay, A.~Puhm and A.~Strominger,
``Conformally Soft Photons and Gravitons,''
JHEP \textbf{01}, 184 (2019)



\bibitem{donnaypastepuhm}
L.~Donnay, S.~Pasterski and A.~Puhm,
``Asymptotic Symmetries and Celestial CFT,''
JHEP \textbf{09}, 176 (2020)







\bibitem{repmode}
F.~M.~Haehl, W.~Reeves and M.~Rozali,
``Reparametrization modes, shadow operators, and quantum chaos in higher-dimensional CFTs,''
JHEP \textbf{11}, 102 (2019)


\bibitem{srmode}
K.~Nguyen and J.~Salzer,
``The effective action of superrotation modes,''
JHEP \textbf{02}, 108 (2021)


\bibitem{pastebaner}
S.~Banerjee and S.~Pasterski,
``Revisiting the shadow stress tensor in celestial CFT,''
JHEP \textbf{04}, 118 (2023)



\bibitem{gianni1}
G.~Boschetti and M.~Campiglia,
``Log translation invariance of log soft gravitational radiation,''
JHEP \textbf{10}, 105 (2025)



\bibitem{Bergmann:1961zz}
P.~G.~Bergmann,
``'Gauge-Invariant' Variables in General Relativity,''
Phys. Rev. \textbf{124}, 274-278 (1961)

\bibitem{aalog}
A. Ashtekar, ``Logarithmic ambiguities in the description of spatial infinity", Found. Phys. \textbf{15}, 419–431 (1985)










  
 
\bibitem{zhibo}
A.~Strominger and A.~Zhiboedov,
``Gravitational Memory, BMS Supertranslations and Soft Theorems,''
JHEP \textbf{01}, 086 (2016)
  


\bibitem{BraginskyThorne}
V.~B.~Braginsky and K.~S.~Thorne,
``Gravitational-wave bursts with memory and experimental prospects,''
Nature \textbf{327}, 123-125 (1987)



\bibitem{thornenl}
K.~S.~Thorne,
``Gravitational-wave bursts with memory: The Christodoulou effect,''
Phys. Rev. D \textbf{45}, no.2, 520-524 (1992)















\bibitem{nichoflan}
{\'E}.~{\'E}.~Flanagan and D.~A.~Nichols,
``Conserved charges of the extended Bondi-Metzner-Sachs algebra,''
Phys. Rev. D \textbf{95}, no.4, 044002 (2017)
[erratum: Phys. Rev. D \textbf{108}, no.6, 069902 (2023)]
doi:10.1103/PhysRevD.95.044002



\bibitem{compnich}
G.~Comp{\`e}re and D.~A.~Nichols,
``Classical and Quantized General-Relativistic Angular Momentum,''
[arXiv:2103.17103 [gr-qc]].


\bibitem{HPS}
S.~W.~Hawking, M.~J.~Perry and A.~Strominger,
``Superrotation Charge and Supertranslation Hair on Black Holes,''
JHEP \textbf{05}, 161 (2017)






\bibitem{BT} 
  G.~Barnich and C.~Troessaert,
  ``Aspects of the BMS/CFT correspondence,''
  JHEP {\bf 1005}, 062 (2010)


\bibitem{bmsw}
L.~Freidel, R.~Oliveri, D.~Pranzetti and S.~Speziale,
``The Weyl BMS group and Einstein{\textquoteright}s equations,''
JHEP \textbf{07}, 170 (2021)



\bibitem{geillzwikpartial}
M.~Geiller and C.~Zwikel,
``The partial Bondi gauge: Further enlarging the asymptotic structure of gravity,''
SciPost Phys. \textbf{13}, 108 (2022)


  \bibitem{geroch} 
  R.~Geroch,
  ``Asymptotic structure of space-time,'' 
 in  \emph{Asymptotic structure of space-time}, ed. L. Witten, Plenum,
New York (1976)






\bibitem{discretebasis}
L.~Freidel, D.~Pranzetti and A.~M.~Raclariu,
``A discrete basis for celestial holography,''
JHEP \textbf{02}, 176 (2024)


\bibitem{crestothesis}
N.~Cresto,
``Asymptotic Higher Spin Symmetries: Noether Realization {\&} Algebraic Structure in Einstein-Yang-Mills Theory,''
[arXiv:2509.17137 [hep-th]].



\bibitem{conformalbasis}
S.~Pasterski and S.~H.~Shao,
``Conformal basis for flat space amplitudes,''
Phys. Rev. D \textbf{96}, no.6, 065022 (2017)








\bibitem{BTprl} 
  G.~Barnich and C.~Troessaert,
  ``Symmetries of asymptotically flat 4 dimensional spacetimes at null infinity revisited,''
  Phys.\ Rev.\ Lett.\  {\bf 105}, 111103 (2010)





\bibitem{raclariulects}
A.~M.~Raclariu,
``Lectures on Celestial Holography,''
[arXiv:2107.02075 [hep-th]].


\bibitem{pastelects}
S.~Pasterski,
``Lectures on celestial amplitudes,''
Eur. Phys. J. C \textbf{81}, no.12, 1062 (2021)


\bibitem{donnayreview}
L.~Donnay,
``Celestial holography: An asymptotic symmetry perspective,''
Phys. Rept. \textbf{1073}, 1-41 (2024)




\bibitem{cgw}
G.~Comp\`ere, S.~E.~Gralla and H.~Wei,
``An asymptotic framework for gravitational scattering,''
Class. Quant. Grav. \textbf{40}, no.20, 205018 (2023)





\bibitem{clmassive}
M.~Campiglia and A.~Laddha,
``Asymptotic symmetries of gravity and soft theorems for massive particles,''
JHEP \textbf{12}, 094 (2015)

\bibitem{berndaviesnohle}
Z.~Bern, S.~Davies and J.~Nohle,
``On Loop Corrections to Subleading Soft Behavior of Gluons and Gravitons,''
Phys. Rev. D \textbf{90}, no.8, 085015 (2014)


\bibitem{hehuangwen}
S.~He, Y.~t.~Huang and C.~Wen,
``Loop Corrections to Soft Theorems in Gauge Theories and Gravity,''
JHEP \textbf{12}, 115 (2014)



\bibitem{1loop2dstress}
T.~He, D.~Kapec, A.~M.~Raclariu and A.~Strominger,
``Loop-Corrected Virasoro Symmetry of 4D Quantum Gravity,''
JHEP \textbf{08}, 050 (2017)



\bibitem{donnaynguyenruzzi}
L.~Donnay, K.~Nguyen and R.~Ruzziconi,
``Loop-corrected subleading soft theorem and the celestial stress tensor,''
JHEP \textbf{09}, 063 (2022)
[erratum: JHEP \textbf{24}, no.2, 116 (2020)]


\bibitem{pasterskiloopcomment}
S.~Pasterski,
``A comment on loop corrections to the celestial stress tensor,''
JHEP \textbf{01}, 025 (2023)


\bibitem{agrawaldonnaynguyenruzzi}
S.~Agrawal, L.~Donnay, K.~Nguyen and R.~Ruzziconi,
``Logarithmic soft graviton theorems from superrotation Ward identities,''
JHEP \textbf{02}, 120 (2024)


\bibitem{kadhe}
S.~Choi, A.~Kadhe and A.~Puhm,
``Long-Range Interactions in Celestial CFT,''
[arXiv:2601.05951 [hep-th]].




\bibitem{oli}
G.~Comp{\`e}re, R.~Oliveri and A.~Seraj,
``The Poincar{\'e} and BMS flux-balance laws with application to binary systems,''
JHEP \textbf{10}, 116 (2020)

\bibitem{coad}
G.~Barnich and R.~Ruzziconi,
``Coadjoint representation of the BMS group on celestial Riemann surfaces,''
JHEP \textbf{06}, 079 (2021)







\bibitem{sayalimatching}
S.~Atul Bhatkar,
``New asymptotic conservation laws for electromagnetism,''
JHEP \textbf{02}, 082 (2021)


\bibitem{fontainematching}
G.~Comp{\`e}re, D.~Fontaine and K.~Nguyen,
``Electromagnetic multipole expansions and the logarithmic soft photon theorem,''
SciPost Phys. Core \textbf{8}, 066 (2025)





\bibitem{aabook}
A.~Ashtekar, “Asymptotic Quantization”, Naples, Italy: Bibliopolis (1987)





\bibitem{logwalok}
M.~Campiglia and A.~Laddha,
``Loop Corrected Soft Photon Theorem as a Ward Identity,''
JHEP \textbf{10}, 287 (2019)


\bibitem{sayalifeynman}
S.~Atul Bhatkar,
``Asymptotic conservation law with Feynman boundary condition,''
Phys. Rev. D \textbf{103}, no.12, 125026 (2021)


\bibitem{KrausQED}
S.~Kim, P.~Kraus, R.~Monten and R.~M.~Myers,
``S-matrix path integral approach to symmetries and soft theorems,''
JHEP \textbf{10}, 036 (2023)



\bibitem{KrausYM}
P.~Kraus and R.~M.~Myers,
``Carrollian partition function for bulk Yang-Mills theory,''
JHEP \textbf{08}, 180 (2025)






\bibitem{bissidonnayvalsesia}
A.~Bissi, L.~Donnay and B.~Valsesia,
``Logarithmic doublets in CCFT,''
JHEP \textbf{12}, 031 (2024)



\bibitem{donnaynguyenruzzilog}
S.~Agrawal, L.~Donnay, K.~Nguyen and R.~Ruzziconi,
``Logarithmic soft graviton theorems from superrotation Ward identities,''
JHEP \textbf{02}, 120 (2024)


\bibitem{chipum2}
S.~Choi, A.~Laddha and A.~Puhm,
``The classical super-rotation infrared triangle. Classical logarithmic soft theorem as conservation law in gravity,''
JHEP \textbf{04}, 138 (2025)




\end{thebibliography}
\end{document}